\documentclass[10pt,aps,twocolumn,superscriptaddress,floatfix,preprintnumbers,prd,nofootinbib]{revtex4-1}
	
\usepackage{array}[=2016-10-06]
    
\usepackage{subcaption}
\usepackage{ragged2e}
\usepackage{amsmath}
\DeclareCaptionJustification{justified}{\justifying}
\makeatletter
\newcommand{\subsetsim}{\mathrel{\mathpalette\subset@sim\relax}}
\newcommand{\subset@sim}[2]{%
  \vtop{\offinterlineskip\m@th
    \ialign{\hfil##\cr
      $#1\subset$\cr\noalign{\kern0.5pt}\scalebox{0.9}{$#1\sim$}\cr
    }%
  }%
}
\makeatother

\usepackage{multirow}
\usepackage{tabularx}
\usepackage{makecell}

\usepackage{amssymb,amsmath,verbatim,mathtools,needspace,enumitem,etoolbox,graphicx,physics,microtype,afterpage,bm}
\usepackage[dvipsnames, usenames]{xcolor}
\usepackage[normalem]{ulem}
\usepackage{booktabs}
\usepackage[unicode, colorlinks=true, linkcolor=linkcolor, citecolor=linkcolor, filecolor=linkcolor,urlcolor=linkcolor, pdfusetitle]{hyperref}
\usepackage[all]{hypcap}
\usepackage[T1]{fontenc}
\usepackage[utf8]{inputenc}
\usepackage{tabularx}
\usepackage{float}
\definecolor{darkpurple}{RGB}{48, 0, 72} 

\usepackage{multirow}
\usepackage{pifont}
\usepackage{lmodern}

\usepackage{orcidlink}

\allowdisplaybreaks
\usepackage{tikz}
\usepackage{color}
\usepackage{framed}
\usepackage{hyperref}
\hypersetup{colorlinks, citecolor=bluscuro, linkcolor=black, urlcolor=bluscuro}
\definecolor{rossos}{cmyk}{0,1,1,0.55}
\definecolor{bluscuro}{rgb}{0.15, 0.2, .85}
\definecolor{bluchiaro}{cmyk}{1,.3,0.,0.1}
\definecolor{ForestGreen}{rgb}{0.13, 0.55, 0.13}

\def\bea{\begin{eqnarray}}
\def\eea{\end{eqnarray}}

\newcommand{\bs}{\begin{subequations}}
\newcommand{\es}{\end{subequations}}

\newcommand{\be}{\begin{equation}}
\newcommand{\ee}{\end{equation}}

\def\lsim{\mathrel{\rlap{\lower4pt\hbox{\hskip0.5pt$\sim$}}
    \raise1pt\hbox{$<$}}}         
\def\gsim{\mathrel{\rlap{\lower4pt\hbox{\hskip0.5pt$\sim$}}
    \raise1pt\hbox{$>$}}}         

\renewcommand{\max}{{\rm max}}

\newcommand{\sapienza}{Dipartimento di Fisica, Sapienza Università 
	di Roma, Piazzale Aldo Moro 5, 00185, Roma, Italy}
\newcommand{\infn}{INFN, Sezione di Roma, Piazzale Aldo Moro 2, 00185, Roma, Italy}
\newcommand{\cern}{Department of Theoretical Physics, CERN, Esplanade des Particules 1, P.O. Box 1211, Geneva 23, Switzerland}
\newcommand{\jhu}{Department of Physics and Astronomy, Johns Hopkins University, 3400 N. Charles
Street, Baltimore, MD 21218, USA}

\newcommand{\unipd}{Dipartimento di Fisica e Astronomia ``G. Galilei'', Università degli Studi di Padova, via Marzolo 8, I-35131 Padova, Italy}
\newcommand{\infnpd}{INFN, Sezione di Padova, via Marzolo 8, I-35131 Padova, Italy}
\begin{document}

\preprint{CERN-TH-2025-240}

\title{
Inferring black hole formation channels in GWTC-4.0 via parametric mass-spin correlations derived from first principles
}

\author{Emanuele Berti$^{\orcidlink{0000-0003-0751-5130}}$}
\email{berti@jhu.edu}
\affiliation{\jhu}

\author{Francesco Crescimbeni$^{\orcidlink{0009-0001-4088-5443}}$}
\email{francesco.crescimbeni@uniroma1.it}
\affiliation{\sapienza}
\affiliation{\infn}

\author{Gabriele Franciolini$^{\orcidlink{0000-0002-6892-9145}}$}
\email{gabriele.franciolini@uniroma1.it}
\affiliation{\unipd}
\affiliation{\infnpd}
\affiliation{\cern}

\author{Simone Mastrogiovanni$^{\orcidlink{0000-0003-1606-4183} }$}
\email{simone.mastrogiovanni@roma1.infn.it}
\affiliation{\infn}

\author{Paolo Pani$^{\orcidlink{0000-0003-4443-1761}}$}
\email{paolo.pani@uniroma1.it}
\affiliation{\sapienza}
\affiliation{\infn}

\author{Grégoire Pierra$^{\orcidlink{0000-0003-3970-7970}}$}
\email{gregoire.pierra@roma1.infn.it}
\affiliation{\infn}

\date{\today}

\begin{abstract}
We investigate the differences between several proposed formation scenarios for binary black holes (BBHs), including isolated stellar evolution, dynamical assembly in dense clusters and AGN disks, and primordial BHs.  
Our approach exploits the predicted spin features of each formation channel, and adopts parameterized models of the predicted correlations between the spin magnitudes (and orientations) and mass, inspired by first principles. 
Using hierarchical Bayesian inference on the recent GWTC-4.0 dataset, we compare these features across all models and assess how well each scenario explains the data.  
We find that the data strongly favor the presence of a positive correlation between mass and spin magnitude, in agreement with previous studies. Furthermore, the hierarchical scenario provides a better fit to the observations, due to the inclusion of second-generation mergers leading to higher spins at larger masses.  
The current dataset is not informative enough about spin orientation: the cluster (random orientations) and AGN (aligned orientations) scenarios have comparable Bayesian evidence.  
Finally, the mass-spin correlation predicted by the primordial scenario gives a poor fit to the data, and this scenario can only account for a subset of the observed events.

\end{abstract}

\maketitle

\tableofcontents
\hypersetup{linkcolor=bluscuro}

\section{Introduction}
\label{sec:introduction}
Since the first gravitational-wave~(GW) detection by the LIGO-Virgo Collaboration~\cite{LIGOScientific:2016aoc}, the growing catalog of GW candidates has provided increasingly precise constraints on the astrophysical population of merging binary black holes (BBHs) and binary neutron stars~\cite{LIGOScientific:2025pvj}.  
With the expansion of the search volume to cosmological distances, it is now possible to investigate the statistical properties of these sources and compare them against different models of stellar and binary evolution, as well as more exotic scenarios.

A central goal of population studies is to determine how compact binaries form and evolve before merger~\cite{Mandel:2018hfr, Mapelli:2018uds, Mapelli:2021taw}. Proposed astrophysical formation channels include for example isolated binary evolution in galactic fields (see e.g.~\cite{Belczynski:2001uc,Hurley:2002rf,Dominik:2012kk,Dominik:2013tma,Dominik:2014yma,Stevenson:2017tfq,COMPASTeam:2021tbl,COMPASTeam:2025fjl,Fragos:2022jik,Andrews:2024saw}), where processes such as mass transfer and common-envelope evolution~\cite{Ivanova:2012vx} shape the final system, and dynamical assembly in dense stellar environments, like globular clusters or galactic nuclei~\cite{Miller:2001ez,Rodriguez:2016kxx,Bartos:2016dgn,Gerosa:2017kvu,Fishbach:2017dwv,DiCarlo:2019pmf,Kimball:2020opk,Kimball:2020qyd,Bouffanais:2021wcr, Afroz:2024fzp, Afroz:2025efn} (see~\cite{Gerosa:2021mno} for a review). These different pathways are expected to leave characteristic imprints on the observed mass, spin, eccentricity and redshift distributions of the merging binaries, which can be probed through hierarchical Bayesian analysis~\cite{Samsing:2017xmd,Stevenson:2017dlk,Baibhav:2020xdf}. 
The growing number of detections enables more detailed tests of stellar evolution predictions, such as the existence of features in the BBH mass distribution associated with pair-instability supernovae~\cite{Farmer:2019jed,Karathanasis:2022rtr,Tong:2025wpz, Afroz:2025ikg}, and the effects of tidal interactions and of the binary's accretion history on the spin distributions~\cite{Gerosa:2018wbw}. Measurements of the merger rate evolution with redshift further connect compact binaries to the cosmic star-formation history. In addition, GW observations offer a unique window onto more exotic possibilities, such as primordial BHs~(PBHs) formed in the early Universe~\cite{Hawking:1974rv,Carr:1975qj,LISACosmologyWorkingGroup:2023njw,Byrnes:2025tji}, which may contribute to the observed merger rate and provide clues about physics beyond the standard model of cosmology, including their potential role as dark matter candidates or tracers of high-redshift phenomena.

One interesting possibility to study and characterize different formation channels is through the spin distribution of merging BBHs~\cite{Vitale:2015tea, Rodriguez:2016vmx, Farr:2017uvj, Farr:2017gtv}. 
The third observing run of the LIGO-Virgo-KAGRA~(LVK) Collaboration~\cite{KAGRA:2021duu} found that most BBH systems are produced with spin magnitudes that have strong support for $\chi\lesssim0.4$, while the distribution of tilt angles $\theta$ seems to prefer systems with spins above the orbital plane. The GW candidates in the GWTC-4.0 catalog~\cite{LIGOScientific:2025pvj} support these conclusions, and also hint at a more detailed structure in the distribution of the effective spin, $\chi_\mathrm{eff}$. In particular, the spin magnitude distribution is concentrated at $\chi \lesssim 0.4$, the spin-tilt distribution may peak away from perfect alignment with the orbital angular momentum, and the $\chi_\mathrm{eff}$ distribution is asymmetric around its peak.
There have been several attempts to constrain the population distributions of GW events by leveraging spin measurements to infer the underlying BH merger channels, both with parametric models ~\cite{Franciolini:2022iaa, Biscoveanu:2022qac,Callister:2022qwb,Baibhav:2022qxm,Li:2022gly,Perigois:2023ihi,Heinzel:2023hlb, Pierra:2024fbl, Alvarez-Lopez:2025ltt, Tong:2025xir, Tiwari:2025oah,Wang:2025nhf,Guo:2024wwv,Szemraj:2025fmm,Li:2025iux} and non-parametric ones ~\cite{Callister:2023tgi, Golomb:2022bon, Rinaldi:2023bbd,Heinzel:2024jlc, Heinzel:2024hva, Rinaldi:2025emt,Guttman:2025jkv,Adamcewicz:2025phm,Sridhar:2025kvi} (see e.g.~\cite{Callister:2024cdx} for an in-depth review of the topic).
Having an accurate inference of BH spins, based on a physically motivated parametric models, can potentially result into more robust constraints on the axion parameter space through the phenomenon of BH superradiance (see e.g. \cite{Arvanitaki:2016qwi,Ng:2019jsx,Ng:2020ruv,Aswathi:2025nxa,Caputo:2025oap}).

In this work, we build upon this approach by modeling the spin distribution and its correlation with mass with physically motivated, simplified models derived from first principles for four of the main formation scenarios: BHs formed in isolation (IBHs), BHs formed hierarchically in clusters (HBHs), BHs formed hierarchically in the disks of active galactic nuclei (AGNs), and primordial BHs (PBHs). These simplified yet insightful models are designed to capture the key features of distinct BBH formation channels, and to provide a more direct physical connection between spin measurements and their astrophysical origins. 

\subsection{Executive summary}
Our main findings are based on the modeling of physical correlations between source masses, spin magnitudes, and tilt angles inspired by the four different formation channels listed before, and detailed in Sec.~\ref{sec:chi_pops} below. Here, for the reader's convenience, we summarize the main results of the paper:

\begin{itemize}    

\item We find strong support for a spin magnitude distribution which broadens at high masses. Within our models, this is closer to hierarchical scenarios (namely, HBHs and AGNs), which include both first and second-generation mergers. However, there is no support either in favor of or against a flat spin direction distribution, compared to spin-angular momentum alignment at small masses. Therefore, we cannot distinguish, among the environments we have considered (clusters or AGN disks), the one in which hierarchical mergers are most likely to occur.

\item Our analysis shows a weak preference for multiple spin populations, although we observe that even a single hierarchical scenario---either HBHs or AGNs---could be able to fully capture the spin distribution and its correlation with the mass observed in the GWTC-4.0 data.

\item The sharp mass-spin correlation predicted by the primordial BH scenario, with efficient cosmological mass-spin evolution, is strongly disfavored as the sole explanation of the GWTC-4.0 dataset.

\item The inferred mass-distribution parameters do not change significantly (i.e., beyond the ${\cal O}(1)\sigma$ level) when performing the inference across the different models, or combinations of models, considered in this work, despite their different predictions for the spin distribution.

\item The information on the merger-rate redshift evolution remains subdominant compared to spin information in the current catalog, and/or the data do not require each subpopulation to follow radically different redshift distributions.

\item For scenarios that provide a poor fit to the data, such as the PBH-only case, the hierarchical likelihood is evaluated in regions of parameter space where the Monte Carlo integrals used for the event posteriors or for the selection effects may be insufficiently sampled. Accurate stability estimators
should be used to ensure proper convergence. 

\end{itemize}

\begin{figure*}[t]
    \centering
\includegraphics[width=1\textwidth]{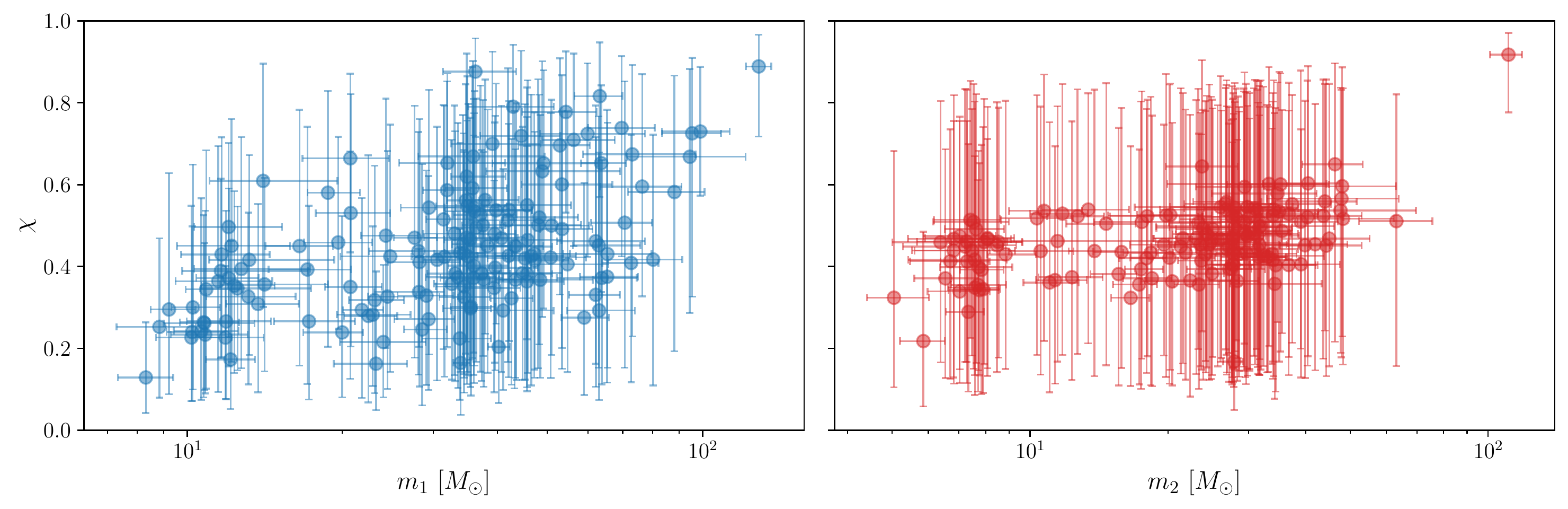}
\includegraphics[width=1\textwidth]{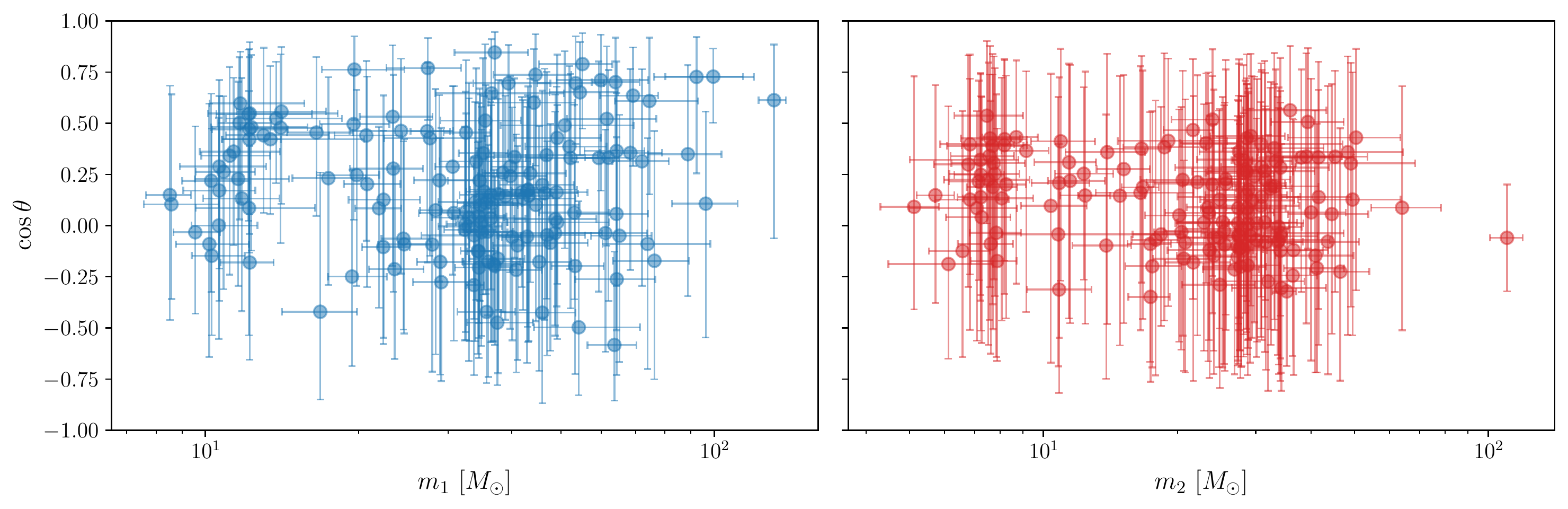}
    \caption{
\textbf{GWTC-4.0 catalog:} Mass-spin scatter plot of the 153 GW candidates used in this analysis, chosen to have at least IFAR=1yr$^{-1}$. 
The stars represent the median values of the BBH event parameters from the GWTC-4.0 catalog.
The x-axis indicates the source-frame masses of the primary (left columns) and secondary (right columns), while the y-axis shows the dimensionless spin magnitudes $\chi$ (first row) or the cosine of the polar angle $\cos \theta$ (second row). Error bars correspond to the $1\sigma$ uncertainties from the official LVK parameter estimation samples for each event. 
We do not show the non-trivial correlation between parameters in the posterior for simplicity. 
$m_i$ indicates source frame mass, as in the rest of the text. 
 }
\label{fig:GWTC-4_data}
\end{figure*}

\section{Modeling different BBH populations}
\label{Pop_models}

In this work, we explore different formation scenarios for BBHs that could explain the physical properties of the population of observed GW sources. We focus on four distinct classes of models:
\begin{itemize}
    \item \textit{IBHs}: binaries formed through isolated stellar evolution in galactic fields.
    \item \textit{HBHs}: binaries assembled dynamically in dense stellar environments, such as globular clusters, where previous merger remnants can participate in subsequent (hierarchical) mergers.
    \item \textit{AGNs}: binaries assembled dynamically in AGN disks. This channel is structurally similar to the HBH one, with the only exception that BHs form in a disk environment, which imprints a preferred aligned-spin direction. 
    \item \textit{PBHs}: binaries formed in the early Universe from the collapse of primordial density fluctuations, independent of stellar processes.
\end{itemize}

Each channel is characterized by its own mass, spin, and redshift distributions, which can be derived from first principles---that is, obtained directly from the physical properties predicted by the channel itself, rather than relying on phenomenological or data-driven assumptions.

While population studies based on distinct mass and redshift distributions in each formation channel remain the primary approach for identifying the nature of the merger populations~\cite{Zevin:2020gbd,Wong:2020ise,Bouffanais:2020qds,Franciolini:2021tla,Colloms:2025hib}, as these properties are more tightly constrained by the GW data, the spin distribution can offer valuable complementary information. Indeed, the most salient spin properties can be modeled agnostically and the prediction of every channel may be more robust, despite the details of the formation mechanism not being fully understood. 
For example, isolated binary formation generically leads to preferentially aligned spins~\cite{Belczynski:2007xg,Gerosa:2013laa,Belczynski:2017gds,Gerosa:2018wbw,Mapelli:2020vfa,Steinle:2020xej,Gangardt:2021lic}, whereas hierarchical mergers predict a subpopulation of spinning BBHs clustering around $\chi \sim 0.7$, which may dominate at large masses~\cite{Berti:2008af,Gerosa:2017kvu,Fishbach:2017dwv,Baibhav:2020xdf}. 
Similar arguments also apply to the primordial scenario. For PBHs, the mass distribution, and even its overall range, is essentially unknown~\cite{Carr:2020gox} and strongly model dependent, whereas the main feature of their spin, namely that PBHs are formed with nearly zero spin in the standard formation scenario~\cite{bbks,DeLuca:2019buf,Mirbabayi:2019uph} and can acquire spin only through mass-dependent accretion~\cite{DeLuca:2020qqa,DeLuca:2020bjf,DeLuca:2023bcr}, is robust and generic. 
In our analysis, our goal is to disentangle the different possible formation channels using only information from the mass-spin correlations, specific to each scenario.
 
In Fig.~\ref{fig:GWTC-4_data} we show the spin magnitudes and orientations of the \textit{observed} (i.e., after imposing the selection effect) primary and secondary source-frame masses of all the GWTC-4.0 BBH  candidates with IFAR=1 yr$^{-1}$, which selects $N_{\rm BBH}$=153 events. Some visible trends can be identified from the figure, such as a positive correlation between mass and spin for the primary BH, a sparsely populated region at negative orientations for low masses, and the outstanding event GW231123~\cite{LIGOScientific:2025rsn}, which is currently the most massive and most rapidly spinning system detected to this day (but see Ref.~\cite{Ray:2025rtt} for a discussion, as the origin of this event is still debated~\cite{Yuan:2025avq,Cuceu:2025fzi,Tanikawa:2025fxw,Croon:2025gol,DeLuca:2025fln,Popa:2025dpz,Li:2025pyo,Paiella:2025qld,Fabj:2025vza,Passenger:2025acb}).

In the following sections, we introduce the phenomenological parametric models for the spin and redshift distributions used in this work. 
We stress that we adopt an agnostic, flexible parametric model for the
merger rate, without imposing astrophysical priors. Leveraging the existing uncertainties on the astrophysical and primordial rates, we assume a logarithmic flat prior on the rate of each population. 
As for the mass spectrum, we assume a single common distribution, i.e., we do not model the mass distribution of each channel separately. Although this is admittedly a strong assumption, it remains compatible with current population-synthesis predictions, which show that isolated, multigenerational, and/or primordial BBHs---either individually or in combination---can reproduce the observed phenomenological mass function. Also, this assumption is analogous to some LVK studies~\cite{LIGOScientific:2025pvj}, where multi-population analyses are performed with a single mass distribution.
The overall mass distribution we adopt, following the LVK analysis, is discussed in Appendix~\ref{app:masspos}. In some cases we will also allow for different merger-rate evolutions for each subpopulation, and find that redshift information remains largely subdominant.

\begin{figure*}[!t]
    \centering
\includegraphics[width=.23\linewidth]{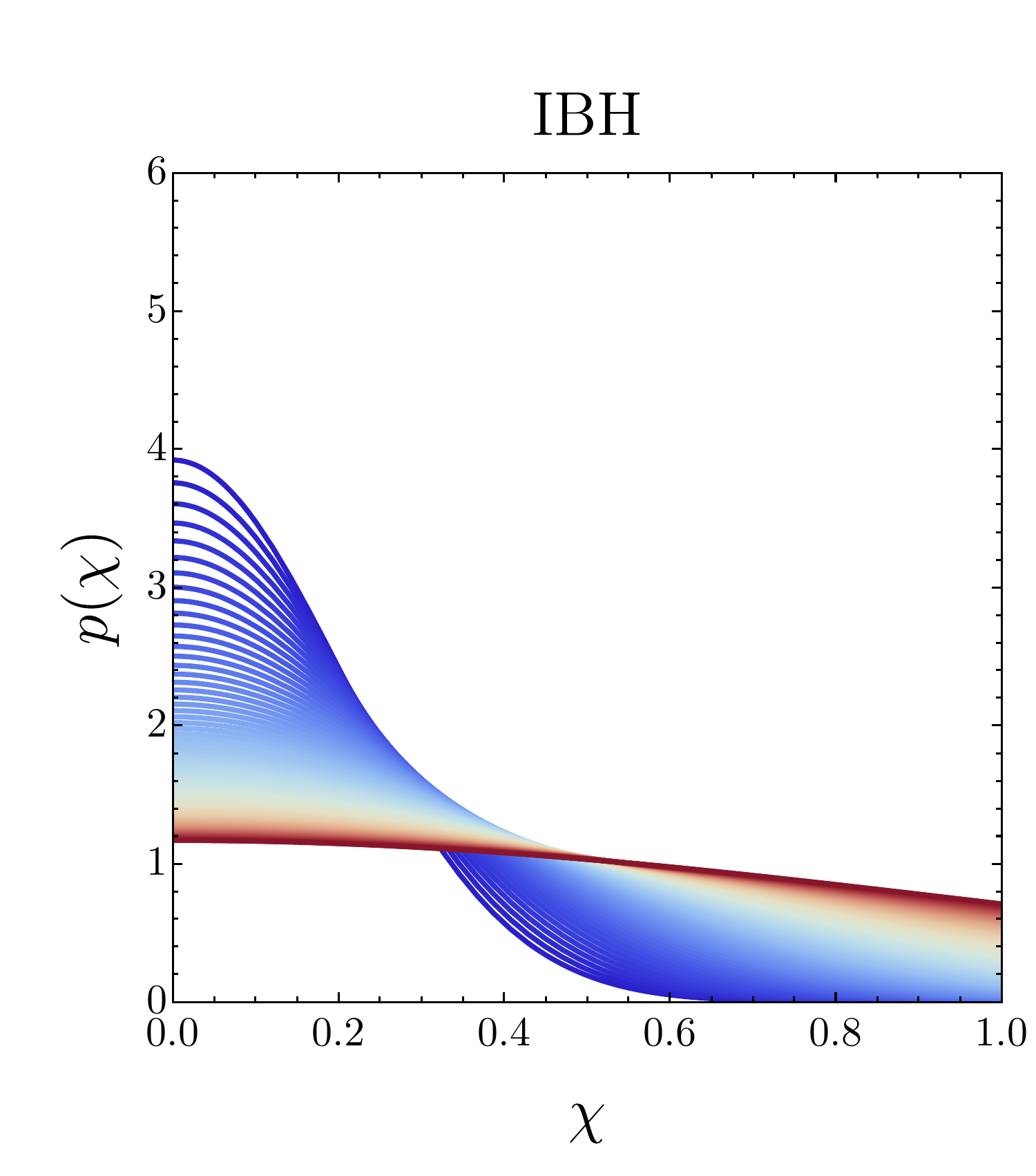}
\includegraphics[width=.23\linewidth]{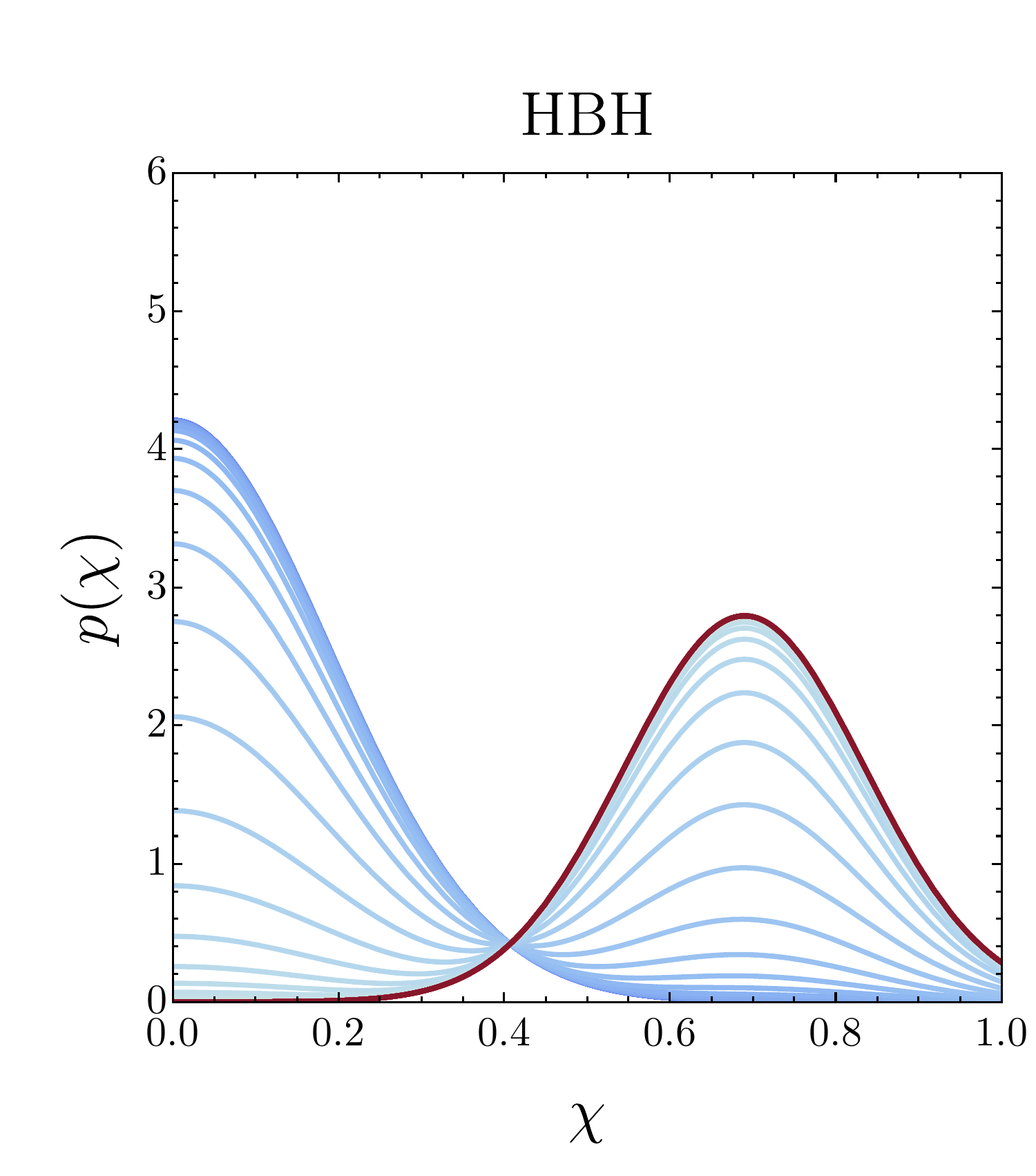}
\includegraphics[width=.23\linewidth]{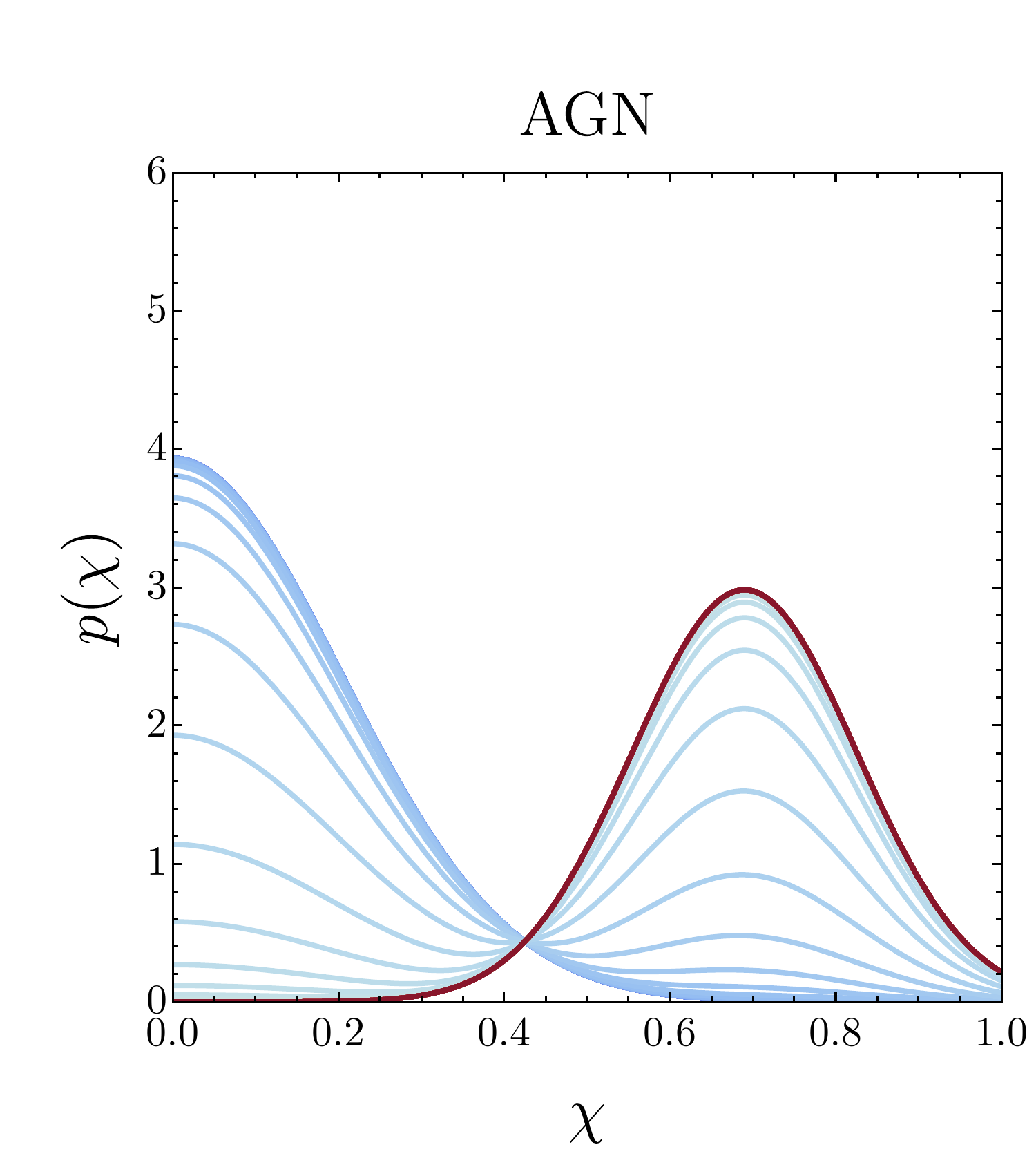}
\includegraphics[width=.28\linewidth]{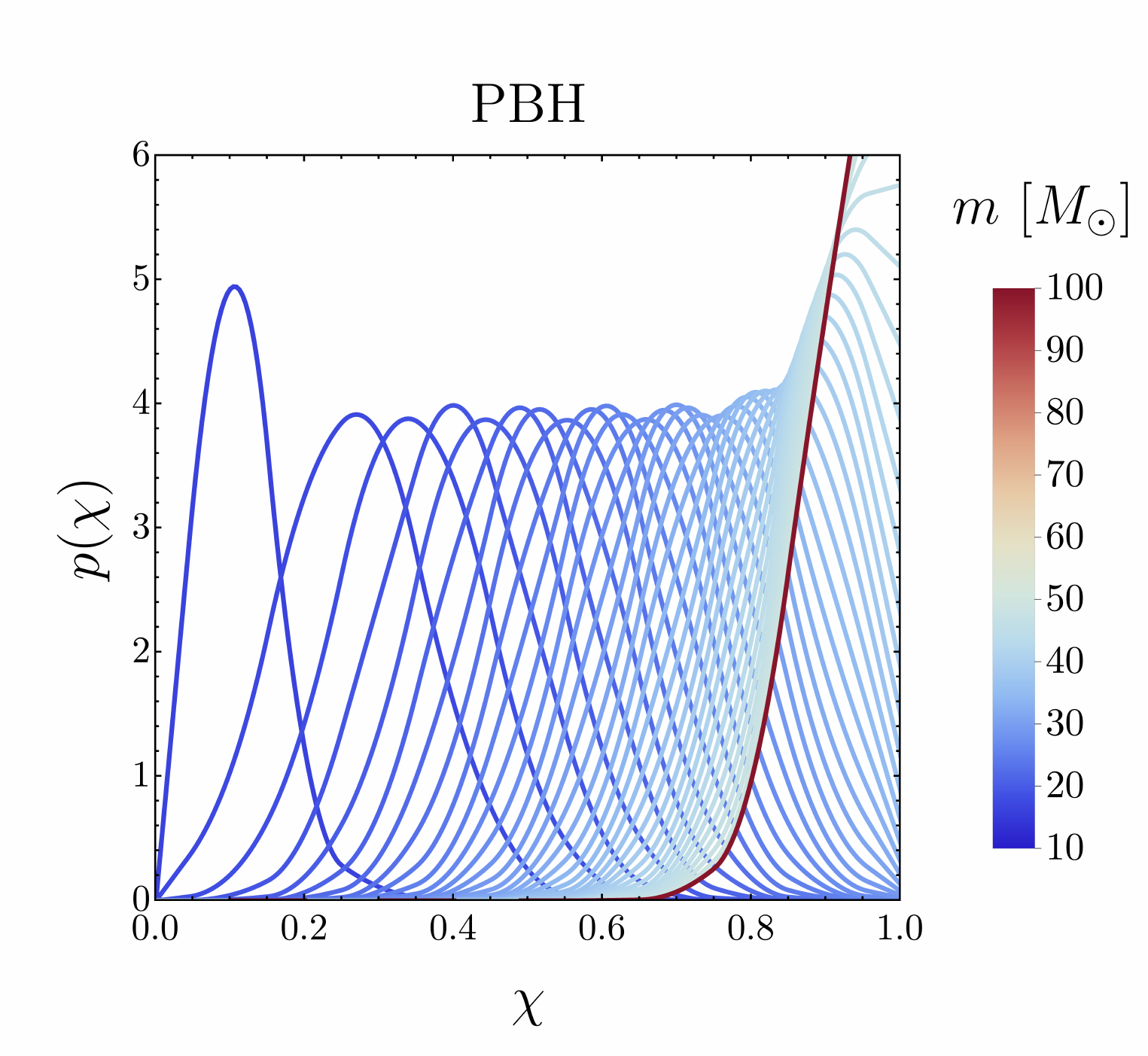}
\includegraphics[width=.23\linewidth]{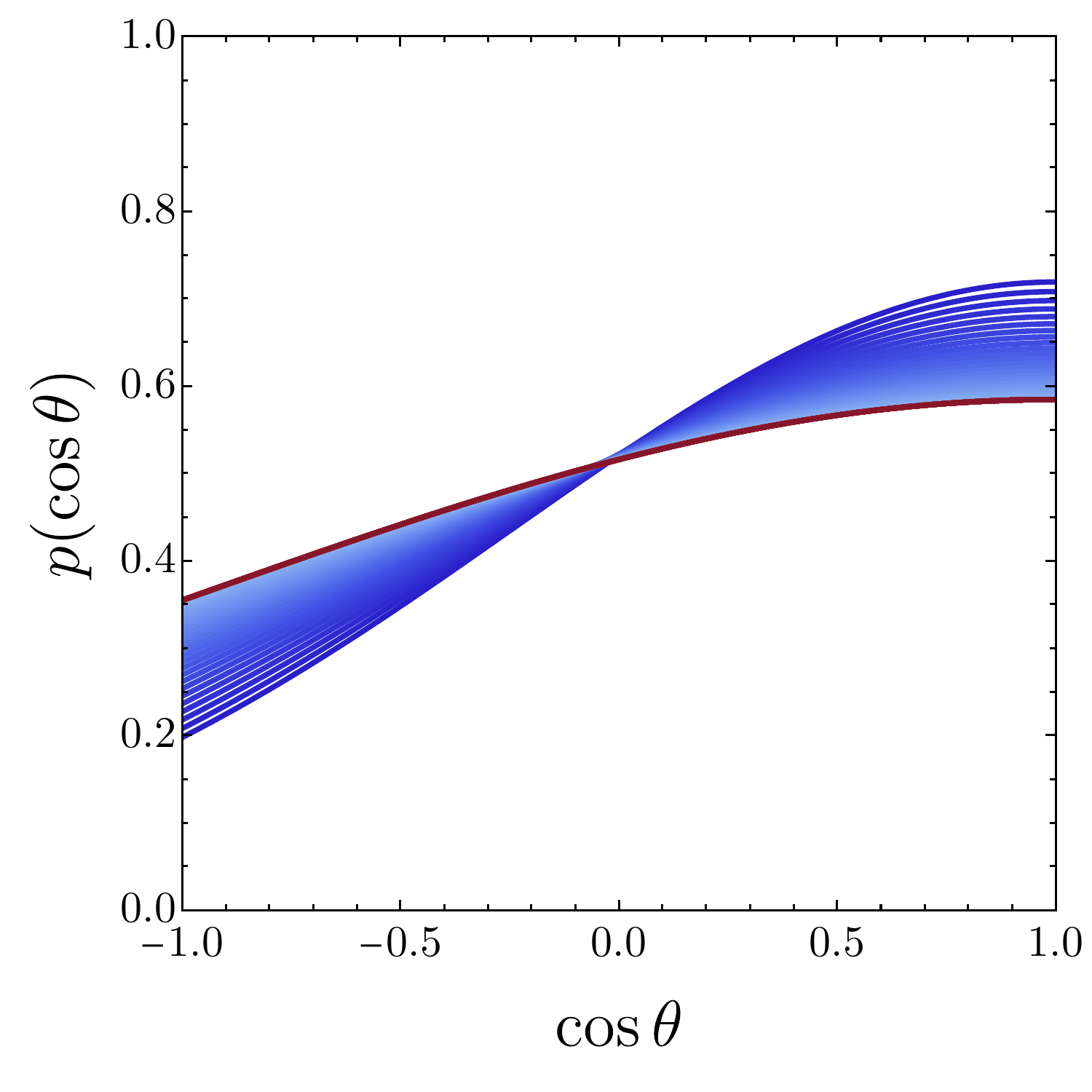}
\includegraphics[width=.23\linewidth]{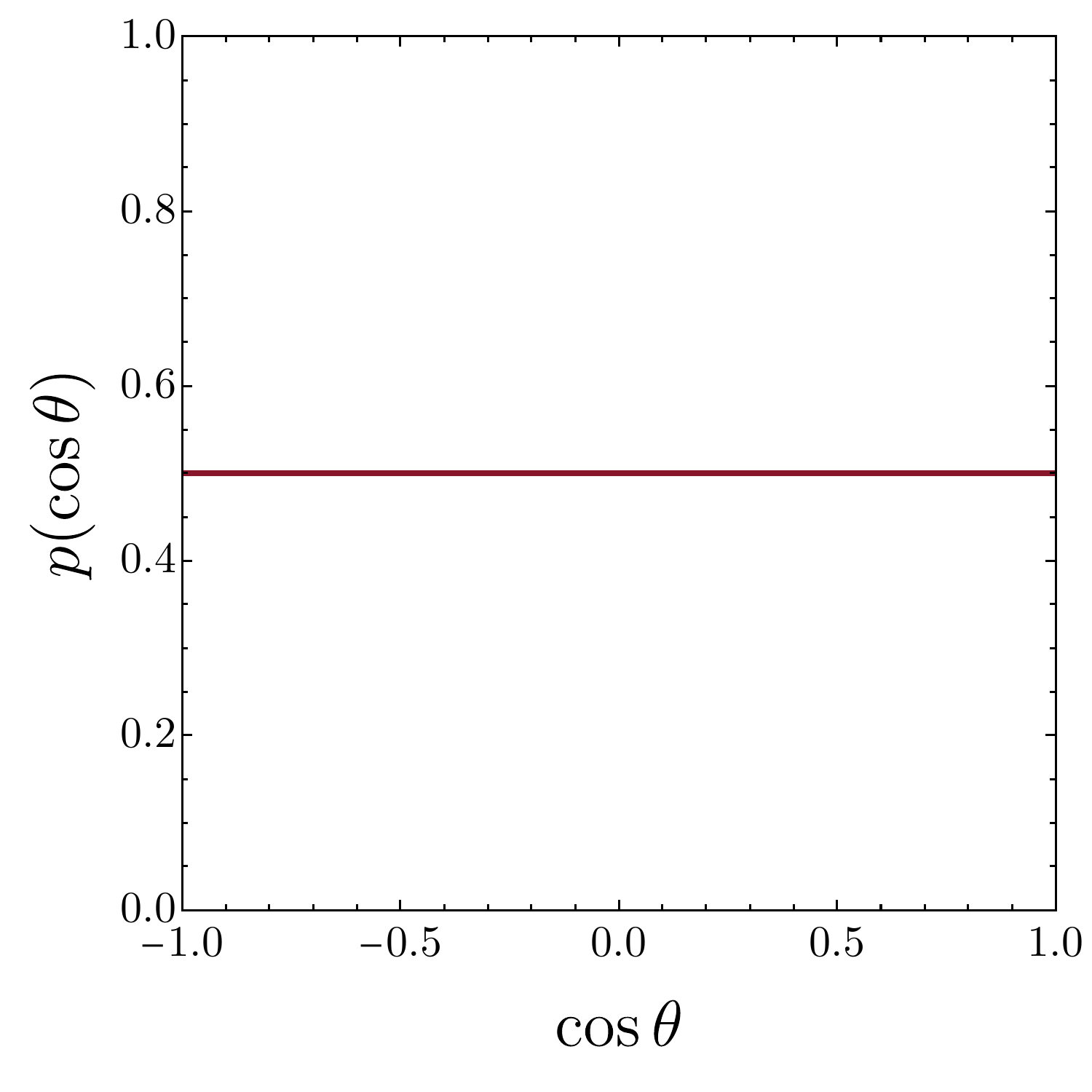}
\includegraphics[width=.23\linewidth]{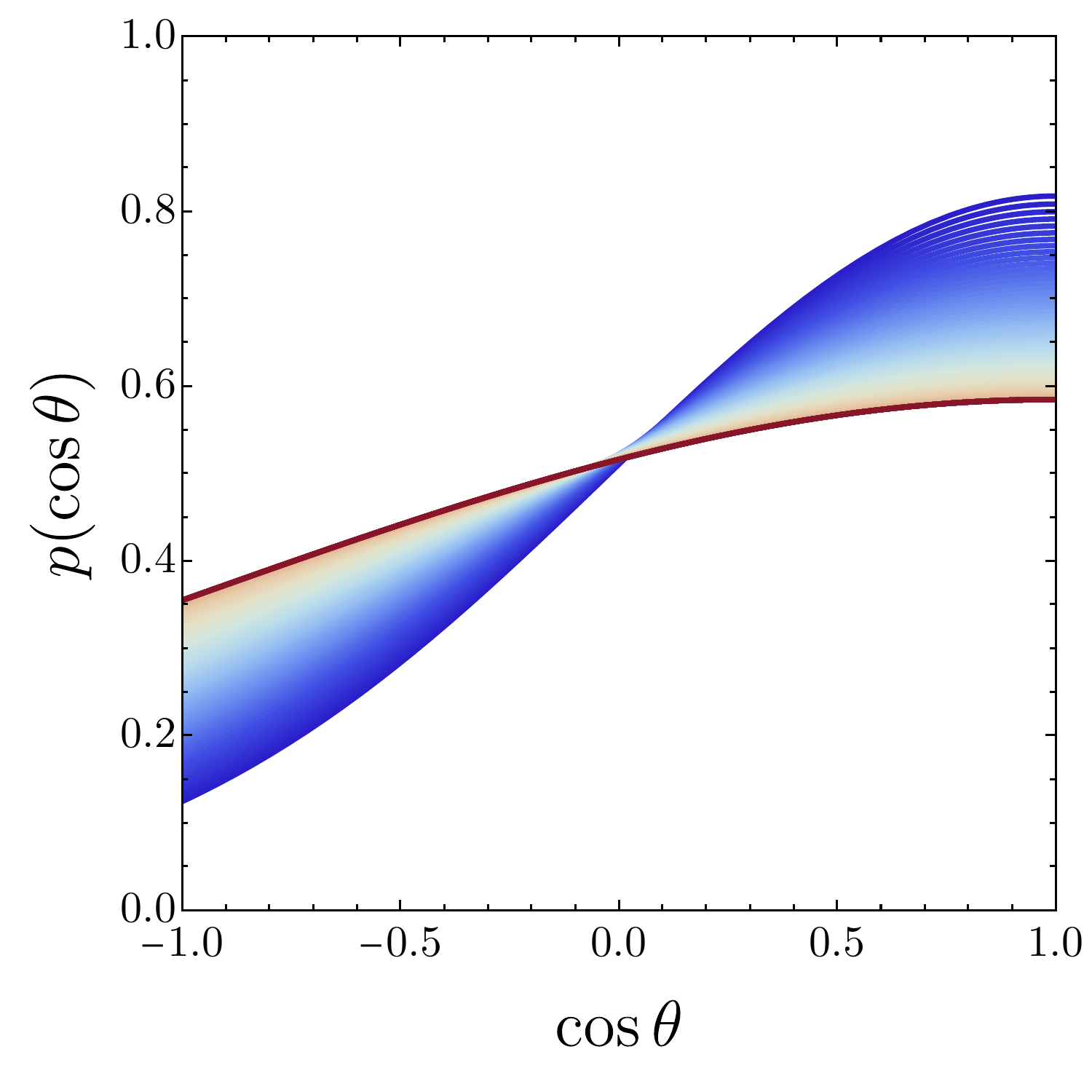}
\includegraphics[width=.28\linewidth]{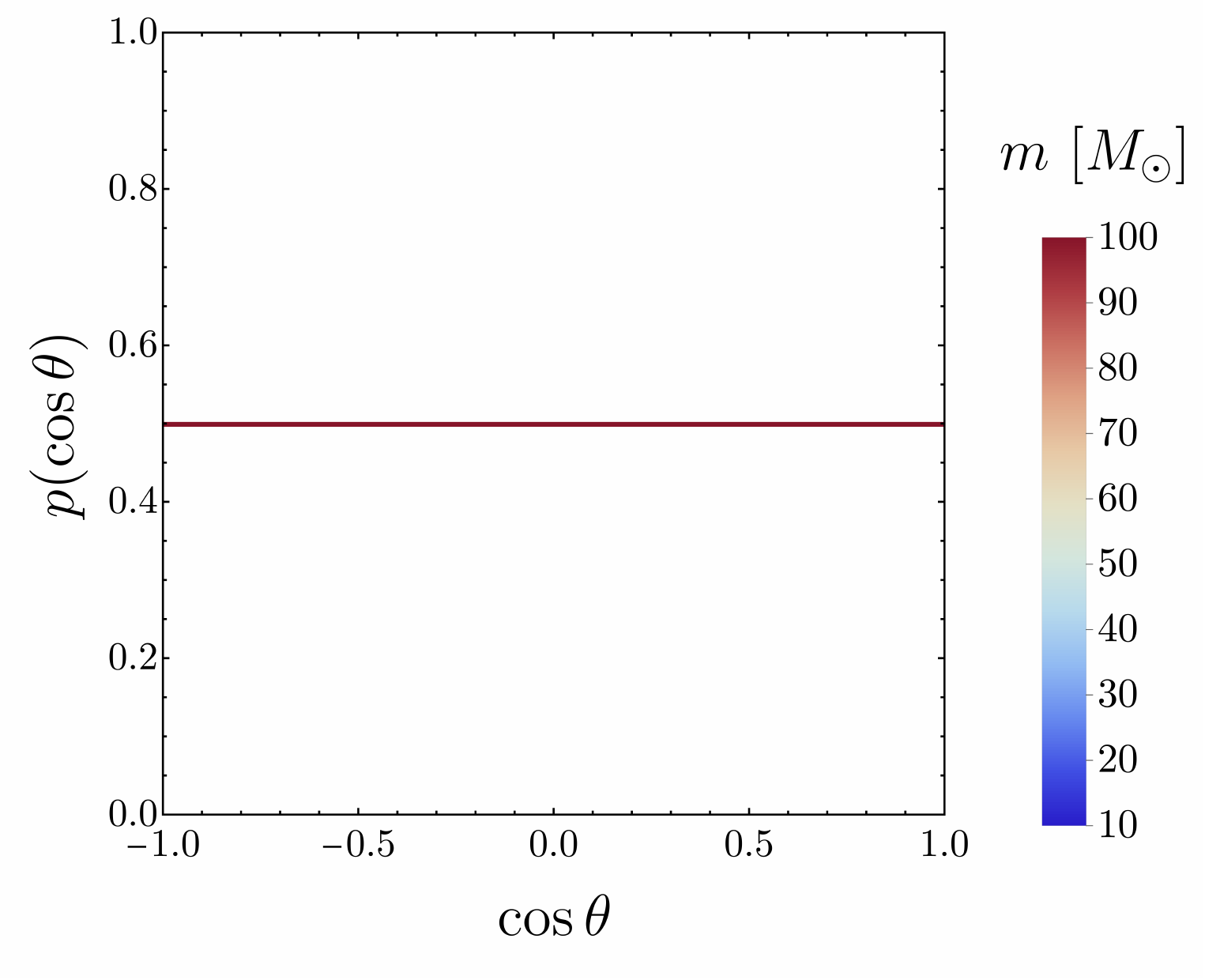}
    \caption{Examples of probability distributions of $\chi$ (first row) and $\cos{\theta}$ (second row), for the different formation channels. The distributions span from 10$M_{\odot}$ (blue) to 100$M_{\odot}$ (red).
    The model parameters of each scenario are fixed to the maximum likelihood (ML) point obtained when fitting the GWTC-4.0 catalog with our models. 
    }
\label{fig:GWTC-4_models}
\end{figure*}

\subsection{Spin models}\label{sec:chi_pops}

In this subsection, we describe in detail the physically-informed models adopted for the spin distributions as functions of the masses. For each population model, we provide an illustrative example of the mass-conditional spin and tilt-angle distributions in Fig.~\ref{fig:GWTC-4_models}. We note that the spin distributions we employ, depending on the specific formation scenario, involve mass-dependent parameters, as detailed below. This introduces characteristic mass-spin correlations in the BBH population.

\subsubsection{Isolated channel}

In isolated binaries, spin orientations retain memory of the stellar progenitors’ evolutionary history~\cite{Belczynski:2007xg,Gerosa:2013laa,Belczynski:2017gds,Gerosa:2018wbw,Mapelli:2020vfa,Steinle:2020xej,Gangardt:2021lic}. While stellar spins are often assumed to be initially aligned with the orbital angular momentum, misalignment can be introduced by supernova kicks imparted at BH formation, which tilt the orbital plane~\cite{Kalogera:1999tq,Vitale:2014mka,Belczynski:2017gds,Gerosa:2018wbw,Bavera:2020inc}. Tidal interactions during the binary’s evolution can partially counteract this effect by realigning spins
\cite{1981A&A....99..126H,Gerosa:2013laa,Steinle:2020xej}. We will remain agnostic about the detailed modeling of these processes and capture the main properties of this scenario as follows.

We assume that BH spins are preferentially aligned with the orbital angular momentum. The tilt angles $\theta_i$ (for $i = 1,2$) are modeled such that $\cos\theta_i$ follows a Gaussian distribution centered at unity, representing a preference for alignment, with a standard deviation parameter $\delta$ controlling the degree of misalignment. The spin magnitudes $\chi_i$ are drawn independently from a Gaussian distribution $\mathcal{N}_{[0,1]}(\chi_i | 0, \chi_{\rm max})$ centered at zero, indicating a preference for small spin values (see e.g.~\cite{Baibhav:2020xdf}), and truncated to the physical range $ [0,1]$.
The resulting distributions are then given by 
\begin{align}
p(\cos\theta_i) &= \mathcal{N}_{[-1,1]}(\cos\theta_i| 1, \delta ),
\\
\label{eq:spin_IBH}
p(\chi_i) &= \mathcal{N}_{[0,1]}(\chi_i| 0, 
\chi_{\rm max}).
\end{align}
The correlation between source mass and spins is encoded in $\chi_{\rm max}$ and $\delta$. 
For simplicity, we model the $\delta(m)$ and $\chi_{\rm max}(m)$ dependence as linear functions within the relevant range (an approximation that should be sufficient at the present level of measurement precision), and take (e.g. \cite{Safarzadeh:2020mlb})
\begin{align}
    \delta &=
    \left [\delta_0^{\rm IBH} + \dot \delta^{\rm IBH}\, \left (\frac{m}{30 M_\odot} \right )  \right]_{[0.1,2]},
    \\
    \chi_{\rm max} &=
    \left [
    \chi^{\rm IBH}_0 
    + \dot \chi^{\rm IBH}\, \left (\frac{m}{30 M_\odot} \right )
    \right] _{[0.1,1]},
\end{align}
where $m$ is the individual source-frame BH mass. We constrain the parameters to be within the physical range $\delta \in [0.1,2]$ and $\chi^{\rm IBH}_{\rm max} \in [0.1,1]$, respectively. 
Hence, in the IBH scenario, the spin model hyperparameters are 
\begin{equation}
\Lambda_{\rm IBH}= 
\{ \delta_0^{\rm IBH}, \dot \delta^{\rm IBH},
\chi^{\rm IBH}_0, \dot \chi^{\rm IBH}\}.
\end{equation}

\subsubsection{Hierarchical channel}\label{sec:HBH_model}

For the dynamical (hierarchical) formation channel, we consider a population which can include contributions from hierarchical mergers. In this model, the BBH population is composed of three distinct sub-populations~\cite{Gerosa:2017kvu,Baibhav:2020xdf}:

\begin{itemize}
    \item \textbf{1g + 1g} (first-generation mergers): both BHs originate from stellar collapse, and later they form a binary by gravitational dynamics (i.e., capture and multi-body interactions).
    \item \textbf{2g + 1g} or \textbf{1g + 2g} (mixed-generation mergers): one component is a remnant from a previous merger.
    \item \textbf{2g + 2g} (second-generation mergers): both components are merger remnants.
\end{itemize}
We neglect $N>2$ generation mergers, as their current detection rate is expected to be subdominant with respect to the former generations. Indeed, since 2g BHs inevitably have large spins, they receive larger merger recoils, and only clusters with very high escape velocity can successfully retain a meaningful fraction of 3g or higher-generation mergers (see e.g.~\cite{Gerosa:2019zmo,Kritos:2022ggc,Kritos:2022non,Santini:2023ukl}).

For all the spin orientations in this scenario, we assume that the angles $\theta_i$ are isotropic, so $\cos\theta_i$ is drawn uniformly in the physically allowed range $[-1, 1]$, and 
\begin{equation}
    p(\cos\theta_i) = \frac{1}{2}, \quad \cos\theta_i \in [-1, 1]. 
\end{equation}
For first-generation binary components, we assume the spin magnitudes $\chi_i$ to be drawn from a Gaussian distribution centered at $\chi =0$ with a variable width $\chi_{\rm max}$. Therefore, similarly to the IBH channel, we assume 
\begin{align}
p_{1g}(\chi) &= \mathcal{N}_{[0,1]}(\chi_i| 0, \chi_{\rm max}).
\end{align}
For binary components of second generation, we assume a spin magnitude drawn from a truncated Gaussian centered at $\bar{\chi}_f$ (which is $\bar{\chi}_f \approx 0.69$ for spinless binaries~\cite{Buonanno:2007sv,Hofmann:2016yih})
\begin{equation}
p_{2g}(\chi) = \mathcal{N}_{[0,1]}(\chi| \bar{\chi}_f, \sigma_f),
\end{equation}
where $\sigma_f$ is a hyperparameter describing the width of the distribution of the remnant spin (which is a relatively narrow distribution around the spinless results, see e.g. Fig.~8 of Ref.~\cite{Borchers:2025sid}), and the Gaussian is truncated to lie within the physical range $[0,1]$. 

To account for all the sub-populations with relative mixing fractions, we model our total spin distribution as 
\begin{equation}
p(\vec{\chi}|\vec{m}) =
\sum_{x,y=1,2}
\pi_x(m_1)\,\pi_y(m_2)\,
p_{xg}(\chi_1)\,p_{yg}(\chi_2),
\label{p_chi_1g2g}
\end{equation}
where $\vec m=(m_1,m_2)$ denotes the component masses, $\vec \chi = (\chi_1,\chi_2)$ the spin magnitudes, and $\vec \theta=(\theta_1,\theta_2)$ the tilt angles. We further define $\pi_x(m)$ as
\begin{equation}
    \pi_x(m) = 
\begin{cases}
f(m) & \text{if } x = 1 \\
1-f(m) & \text{if } x = 2
\end{cases},
\label{eq:pchiHBH}
\end{equation}
with 
\begin{equation}
    f(m) = f_{\rm 1g,high} + \frac{(f_{\rm 1g,low} - f_{\rm 1g,high})}{1 + e^{\frac{m-m_t}{\delta m_t}}},
\end{equation}
where $f(m)$ is a sigmoid mixture fraction that switches from different fractions of 1g/2g BHs as a function of mass. As 2g BHs should be typically more massive than 1g ones, we expect the fraction $f(m)$ to transition between 1 and 0. 
In the above equation, $f_{\rm 1g,high}$ is the fraction of 1g BHs at high masses, $f_{\rm 1g,low}$ is the fraction of 1g BHs at low masses, $m_t$ is a transition mass, and $\delta m_t$ is a transition window.
Note that Eq.~\eqref{eq:pchiHBH} allows for the possibility that the more massive black hole is 1g and the secondary 2g.

This model of the spin distribution of 1g+2g BHs can reproduce the $\chi_{\rm eff}$ distribution found in population-synthesis results for GC and NSC environments, as defined in Ref.~\cite{Zevin:2020gbd} based on \cite{Antonini:2018auk,Rodriguez:2019huv}, simply by adjusting $\chi_{\rm max}$ and the relative fraction as a function of the primary mass. This was shown explicitly in Ref.~\cite{Franciolini:2022iaa} (see their Fig.~1). Therefore, this model provides a well-motivated framework to capture the spin properties, and their correlation with mass, in the dynamical scenario. 

Having fixed $\{ \bar{\chi}_f^{\rm HBH} =0.69,
f_{\rm 1g,low}^{\rm HBH} = 1, 
f_{\rm 1g,high}^{\rm HBH} = 0 \}$
to their values as motivated above, the population hyperparameters in the HBH model are
\begin{equation}
\Lambda_{\mathrm{HBH}} =
\{ \chi_{\mathrm{max}}^{\rm HBH}, 
\sigma_f^{\rm HBH}, 
 m_t^{\rm HBH}, \delta m_t ^{\rm HBH} \}.
\end{equation}

\subsubsection{Active galactic nuclei channel}

We also include a simple model for the AGN channel (see e.g.~\cite{Bartos:2016dgn,Cook:2024ajp}). For the spin magnitude, this channel is assumed to contain both 1g and 2g mergers, with a spin distribution matching the form adopted for the HBH model in Sec.~\ref{sec:HBH_model}. However, for this model, the presence of a disk in the AGN environment generates a preferential direction for both BH pair dynamics and gas accretion~\cite{Bardeen:1972fi, Bardeen:1975zz}, thus leading to binaries that have nearly aligned spins~\cite{Santini:2023ukl,Cook:2024ajp}. 
Therefore, we model the tilt distributions as:
\begin{equation}
    p(\cos\theta_i) = \mathcal{N}_{[-1,1]}(\cos\theta_i| 1, \delta^{\rm AGN} ).
\end{equation}
We allow for the tilt angles to be correlated with masses, therefore:
\begin{align}
    \delta^{\rm AGN} &=
    \left [\delta_0^{\rm AGN} + \dot \delta^{\rm AGN}\, \left (\frac{m}{30 M_\odot} \right )  \right]_{[0.1,2]}.
\end{align}
Following a similar treatment for the spin magnitude as described above, the AGN model is described by the following model hyperparameters:
\begin{equation}
\Lambda_{\mathrm{AGN}} = \{ 
\delta_0^{\rm AGN}, \dot \delta^{\rm AGN}, \chi_{\mathrm{max}}^{\rm AGN}, 
 \sigma_f^{\rm AGN}, 
 m_t^{\rm AGN},
 \delta m_t^{\rm AGN} \}.
\end{equation}
Accretion effects in the AGN channel are not explicitly included in this model, and are left for future work.

\subsubsection{Primordial channel}

In the PBH scenario, binaries are supposed to form in the early Universe with negligible initial spins, due to the quasi-spherical nature of the collapse of large curvature perturbations during radiation domination~\cite{bbks,DeLuca:2019buf,Mirbabayi:2019uph}. However, PBHs may acquire spin through gas accretion before re-ionization~\cite{DeLuca:2020qqa,DeLuca:2020bjf}, with the efficiency of accretion depending on the PBH mass. Accretion is negligible for light PBHs below ${\cal O}(10) M_\odot$, while it can induce significant spins for more massive PBHs (see the upper right panel of Fig.~\ref{fig:GWTC-4_models}), thus introducing a characteristic correlation between spin and mass. The location of this transition depends on the accretion efficiency, and it is encoded in the model hyperparameter $z_{\rm cut-off}$~\cite{DeLuca:2020bjf}.
In the absence of subsolar mass merger detections \cite{LVK:2022ydq,Nitz:2022ltl,Crescimbeni:2024cwh,Golomb:2024mmt,Crescimbeni:2024qrq} and access to high redshfit events with $z\gtrsim {\cal O}(30)$ \cite{Nakamura:2016hna,Koushiappas:2017kqm,DeLuca:2021hde,Pujolas:2021yaw,Ng:2022agi,Ng:2022vbz,Franciolini:2023opt}, the mass-spin correlation induced by accretion remains the only predictive testable imprint of the primordial scenario \cite{Franciolini:2021xbq}.

The spin directions are assumed to be isotropically distributed, as expected for independent PBHs with random relative orientations~\cite{DeLuca:2020qqa}:
\begin{equation}
p(\cos\theta_i) = \frac{1}{2}, \quad \cos\theta_i \in [-1, 1].
\end{equation}
The spin magnitudes $\chi_i$ are modeled via an analytical fit describing accretion-driven spin growth as a function of the primary mass $m_1$, mass ratio $q$, and a cutoff redshift $z_\text{cut-off}$ that parametrizes the end of efficient accretion.
A detailed description of these functions can be found in Ref.~\cite{Franciolini:2021xbq}. 
Given the uncertainties in the accretion efficiencies, as well as expected scattering in the environmental properties around PBH binaries, we allow both spins to be distributed as a Gaussian centered around the value $\chi = \chi_{\rm PBH}(\vec m|z_\text{cut-off})$ predicted in the model, so that 
\begin{equation}
p(\chi|\vec m) = 
\mathcal{N}_{[0,1]}(\chi\,|\, \chi_{\rm PBH}(\vec{m}| z_{\rm cut-off}), \sigma_\chi).
\end{equation}
In this case, the hyperparameters of the PBH model are
\begin{equation}
\Lambda_{\rm PBH} = \{ z_\text{cut-off}, \sigma_\chi \},
\end{equation}
as the full spin distribution is determined by these parameters via the analytical fit described above.
We summarize the spin model parameters in Table~\ref{tab:priors}.

\begin{table}[t!]
\centering
\caption{Priors on the BBH spin and redshift-evolution model hyperparameters. ${\cal U}$ indicates uniform distribution, while  ${\cal LU}$ indicates log-uniform distribution, within the indicated range. }
\label{tab:priors}
\begin{tabularx}{\linewidth}{|X|l|l|}
\hline\hline
Spin models & Parameter & Prior\\
\hline\hline
\multirow{4}{*}{IBH
} 
  & $\delta_0^{\rm IBH}$ & ${\cal U}[0.1, 1]$ \\
  & $\dot{\delta}^{\rm IBH}$ & ${\cal U}[-1,1]$ \\
  & $\chi_0^{\rm IBH}$ & ${\cal U}[0.05, 1]$ \\
  & $\dot{\chi}^{\rm IBH}$ & ${\cal U}[-1,1]$ \\
\hline
\multirow{4}{*}{HBH} 
  & $\chi_{\max}^{\rm HBH}$ & ${\cal U}[0.1, 1]$ \\
  & $\sigma_f^{\rm HBH}$ & ${\cal U}[0.1,0.5]$ \\
  & $m_t^{\rm HBH}$ & ${\cal U}[10, 100]\,M_\odot$ \\
  & $\delta m_t^{\rm HBH}$ & ${\cal U}[1, 100]\,M_\odot$ \\
\hline
\multirow{6}{*}{AGN} 
 & $\delta_0^{\rm AGN}$ & ${\cal U}[0.1, 1]$ \\
  & $\dot{\delta}^{\rm AGN}$ & ${\cal U}[-1,1]$ \\
  & $\chi_{\max}^{\rm AGN}$ & ${\cal U}[0.1, 1]$ \\
  & $\sigma_f^{\rm AGN}$ & ${\cal U}[0.1,0.5]$ \\
  & $m_t^{\rm AGN}$ & ${\cal U}[10, 100]\,M_\odot$ \\
  & $\delta m_t^{\rm AGN}$ & ${\cal U}[1, 100]\,M_\odot$ \\
\hline
\multirow{2}{*}{PBH} 
  & $z_{\rm cut-off}$ & ${\cal U}[10, 30]$ \\
  & $\sigma_\chi$ & ${\cal U}[0.05, 0.2]$ \\
\hline\hline
Redshift evolution & Parameter & Prior range \\
\hline\hline
\multirow{3}{*}{IBH, HBH, AGN} 
 & $\gamma$ & ${\cal U}[0, 10]$ \\
 & $k$ & ${\cal U}[0, 10]$ \\
 & $z_{\rm p}$ & ${\cal U}[0, 10]$ \\
\hline
\multirow{2}{*}{PBH} 
 & $\gamma_{\rm PBH}$ & $1.17$ \\
 & $k_{\rm PBH}$ & $-\gamma_{\rm PBH}$ \\
 \hline
 \multirow{1}{*}{All models}
  & $R_0^c$ & ${\cal L}{\cal U}[10^{-4},\,80]$ \\
\hline\hline
\end{tabularx}
\end{table}

\subsection{Redshift model}

For all channels, we model the merger rate density evolution with redshift through a smooth Madau-Dickinson broken power-law function,
\begin{equation}\label{eq:redshift_model}
    \psi \left(z | \Lambda_z\right) = \left[1+\frac{1}{\left(1+z_{\mathrm{p}}\right)^{\gamma+k}}\right] \frac{(1+z)^\gamma}{1+\left[
    \frac{(1+z)}{\left(1+z_{\mathrm{p}}\right)} \right]^{\gamma+k}} \, ,
\end{equation}
governed by three parameters 
\begin{equation}
     \Lambda_z \equiv \{ \gamma, k, z_{\mathrm{p}} \}.
\end{equation}
This shape is motivated by the star formation rate evolution~\cite{Madau:2014bja}, although it remains sufficiently flexible, so that it can fit the low-redshift merger rate at $z\lesssim \mathcal{O}(3)$---of relevance for current LVK sensitivity---for all the formation channels considered here. 

Interestingly, the functional form in Eq.~\eqref{eq:redshift_model} can reproduce specific population-synthesis predictions for the scenarios described above.
In particular, as shown in Ref.~\cite{Ng:2020qpk}, the isolated channel predicts a merger rate with reference parameters~\cite{Belczynski:2016obo} 
\begin{align}
\Lambda_z^{\rm IBH}
= \{ \gamma = 2.57 \,,
k = 3.26 \,, 
z_{\mathrm{p}} = 2.36
\}.
\end{align}
For the case of HBHs, the merger rate evolution shown in population synthesis studies (see e.g.~\cite{Rodriguez:2018rmd}) is fitted using 
\begin{align}
\Lambda_z^{\rm HBH}
= \{ \gamma = 1.56 \,,
k =  1.94 \,, 
z_{\mathrm{p}} =2.12
\}.
\end{align}
The PBH merger rate is dominated by the binaries formed at high redshift, before matter-radiation equality, and the merger rate evolution is predicted to be of the form $\psi \propto t^{-34/37}$, where $t$ is the age of the Universe at redshift $z$ (see Ref.~\cite{Raidal:2024bmm} for a recent review). This robust prediction results from the properties of binaries at high redshift and GW-driven evolution through Peter's formula~\cite{Peters:1963ux,Peters:1964zz}.
We can fit this relation at low redshift with better than a few percent accuracy by choosing
\begin{align}
\Lambda_z^{\rm PBH}
= \{ \gamma = -k  = 1.17
\};
\end{align}
this yields a simplified single power-law expression, where the parameter $z_p$ drops out.

\section{Hierarchical Bayesian inference set-up}
\label{sec:total_rate}

In the following, we describe the setup of the analysis that we perform to  identify and disentangle the various channels. We perform a hierarchical Bayesian analysis of the GWTC-4.0 catalog ~\cite{LIGOScientific:2025hdt,LIGOScientific:2025slb} with \texttt{icarogw}~\cite{Mastrogiovanni:2023zbw}, a \texttt{Python} code developed to infer astrophysical and cosmological population properties of noisy, heterogeneous, and incomplete observations. We assume standard cosmological parameters for the $\Lambda$CDM model~\cite{Planck:2015fie}.
The core of the hierarchical Bayesian analysis is the construction of the merger rate, for which we adopt two models:

\begin{itemize}[leftmargin=*]
\item \texttt{Model~I} (common mass and redshift distributions). In this minimal model, both the mass distribution and the redshift evolution are shared across all channels. Hence, only the spin distributions are allowed to vary depending on the formation channel:
    \begin{equation}
    \begin{aligned}
    \frac{d N_{\rm BBH}(\Lambda )}{d\lambda\, dz\, dt} 
    &=
\frac{dV_c}{dz}\,
\frac{\psi(z| \Lambda_z)}{1+z} 
    p (\vec{m} |  \Lambda_m)
    \\
    & \times 
    \sum_{c} R_0^c 
    p_c(\vec{\chi}, \cos\vec{\theta} | \vec m, \Lambda_c),
    \end{aligned}
    \label{eq:modelI_L}
    \end{equation}
    where $c$ takes values in the subset of models $\{ \rm IBH, HBH, AGN, PBH \}$.

\item \texttt{Model~II} (common mass distribution). All the aforementioned channels share a common mass distribution, while both the redshift evolution and spin distributions remain channel-dependent. The overall merger rate is then defined as 
\begin{equation}
\begin{aligned}
    \frac{d N_{\rm BBH}(\Lambda )}{d\lambda\, dz\, dt} 
    &= 
    \frac{dV_c}{dz} 
   \frac{1}{1+z} 
    p (\vec{m} | \Lambda_m)
 \nonumber \\
&\times 
\sum_{c} R_0^c 
\psi(z|\Lambda_z^c) 
p_c(\vec{\chi}, \cos\vec{\theta} | \vec m, \Lambda_c).
\end{aligned}
\label{eq:modelII_L}
\end{equation}
\end{itemize}

Here, $\lambda=(\vec m, \vec \chi, \vec \theta)$ denotes the set of intrinsic binary parameters; $z$ is the redshift; $t$ is the source-frame time; and $\Lambda$ are the model hyperparameters. The term $dV_c/dz$ is the differential comoving volume element, and $R(z|\Lambda)$ describes the redshift-dependent merger rate density. 
The function $p(\vec{m} | \Lambda)$ models the distribution of source-frame component masses, while $p(\vec{\chi}, \cos\vec{\theta} | \vec m, \Lambda)$ specifies the joint distribution between spin magnitudes and spin orientations.
Next, the hierarchical likelihood for $N_{\rm obs}$ GW observations $\{x\}$ can be written as~\cite{Mandel:2018mve}
\begin{equation}
    \begin{aligned}
\mathcal{L}(\{x\}|\Lambda) 
\propto & 
e^{-N_{\exp }(\Lambda)} \prod_{i}^{N_{\mathrm{obs}}} 
T_{\mathrm{obs}} 
\int d\lambda d z 
\mathcal{L}_{\mathrm{obs}}\left(x_{i}| \lambda, z\right)
\\
& \times \frac{d N_{\mathrm{BBH}}(\Lambda)}{d\lambda d z d t},
\end{aligned}
\end{equation}
where $\mathcal{L}_{\mathrm{obs}}\left(x_{i} | \lambda, z\right)$ is the likelihood of the single GW event $x_{i}$. The factor $N_{\rm exp}$, corresponding to the expected number of detectable events for the model with hyperparameters $\Lambda$, encodes the selection effects, and is defined as \begin{equation}
N_{\text{exp}}(\Lambda) = T_{\text{obs}} 
\int d\lambda \, dz \, P_{\text{det}}(\lambda, z) \, \, \frac{dN_{\text{BBH}}(\Lambda)}{dz \, d\lambda \, dt},
\end{equation}
where $T_{\text{obs}}$ is the observation time, and $P_{\text{det}}(\lambda, z)$ denotes the probability of detecting an event with intrinsic parameters $\lambda$ at redshift $z$. An event is considered detected if it exceeds the threshold defined by the search pipeline, such as a minimum signal-to-noise ratio (SNR) or a false-alarm rate (FAR) limit ~\cite{LIGOScientific:2025hdt,LIGOScientific:2025slb}
. Following the LVK state-of-the-art analyses, we impose a cut on the FAR for each event being smaller than ${\rm (FAR_{\rm min})^{-1}} \geq1 $ yr, where ${\rm FAR_{\rm min}}$ is the minimum FAR computed among the search pipelines active at the detection of the specific event. The integrals in the hierarchical likelihood are computed numerically using Monte Carlo (MC) integration of a set of finite samples. To ensure numerical stability and good estimates of these integrals, we use a variance cut~\cite {Talbot:2023pex}. For more details on the numerical likelihood evaluation, see Appendix~\ref{app:likelihood-cuts}.

It has been shown that GW population inference can be biased 
unless the variance of the log-likelihood estimator is below unity, especially when including spin information in the models~\cite{Heinzel:2025ogf}. 
Therefore, following the recent LVK population analysis~\cite{LIGOScientific:2025pvj}, we adopt a threshold of $\sigma^2_{\ln \hat{\mathcal{L}}}=1$ to mitigate potential biases in the posterior. Above this threshold, the likelihood estimate may not be sufficiently converged, and posterior samples with larger variances are therefore discarded. In practice, this can exclude substantial regions of the hyperparameter space for some models, limiting the range of populations that can be robustly explored~\cite{Wysocki:2018mpo,Essick:2022ojx,Doctor:2019ruh,Delfavero:2021qsc,Golomb:2021tll,Mould:2023eca,Hussain:2024qzl,Mancarella:2025uat}.
As shown in Appendix~\ref{app:likelihood-cuts}, the likelihood-variance cut can indeed impact the inference, in particular for models which include spin and/or spiky distributions. When a model fits the data poorly, evaluating the likelihood requires sampling points in the far tail of the posterior of some events, which suffer from poor coverage due to the finite sampling of the GW event posterior, or evaluating the model distribution on regions of parameter space which have few injections needed to compute the selection function (see, e.g., Appendix~D3 of~\cite{LIGOScientific:2025pvj} and Appendix~A of~\cite{Mancarella:2025uat}).

The population models introduced above depend on a set of hyperparameters that describe the underlying distributions of BBH properties. To carry out hierarchical inference, we impose prior distributions on these hyperparameters, as shown in Table~\ref{tab:priors}.
The priors adopted for the mass distribution are shown in Appendix~\ref{app:masspos}.

\section{Results}\label{sec:pop1}

This section presents the results of our inference analysis with the GWTC-4.0 catalog ~\cite{LIGOScientific:2025hdt,LIGOScientific:2025slb}. We begin by assessing which of the formation channels considered (IBH, HBH, AGN, or PBH) are required to reproduce the observed BBH population, first examining each channel individually, then mixtures of two, and finally three-channel configurations. We then discuss, within these combined IBH+HBH+PBH scenarios, how the different channels can jointly contribute to the observed mass and spin distributions.\footnote{The HBH and AGN channels, which differ in our modeling only through the spin orientation, are highly similar and degenerate, as the current dataset does not contain enough information to distinguish between different spin-orientation distributions. For this reason, in the three-channel case we retain the HBH channel as our reference model.}
In the 1- and 2-population cases, we use \texttt{Model I} only, in which the merger rate evolution is shared across all channels, while for the 3-population case we also explore \texttt{Model II} (see Sec.~\ref{sec:total_rate}).

For all the cases listed above, the reconstructed mass distribution is in agreement with the one reconstructed by the LVK Collaboration in Ref.~\cite{LIGOScientific:2025pvj}. In other words, varying the  assumptions on the spin model does not significantly affect the inference of the mass distribution. We discuss this in more detail in Appendix~\ref{app:masspos}.

\begin{table}[t]
\centering
\begin{tabular}{|l|c|c|c|}
\hline
\textbf{Model $\mathcal{M}$} 
& 
$\log_{10}(\mathcal{B}^{\mathcal{M}}_{\bigstar})$  &
${\cal R}^{\cal L}_{\rm max}$ &
${\cal R}^{\cal L}_{\rm av} $
\\
\hline
\hline
\multicolumn{4}{|c|}{1 population}\\
\hline
IBH & 5.3 & 4.7 & 4.9  \\
\hline
HBH & 6.2 & 5.7 & 5.8\\
\hline
AGN & 7.0 & 6.6 & 6.9 \\
\hline
PBH & -- & -- &  -- \\
\hline
\hline
\multicolumn{4}{|c|}{2 populations}\\
\hline

IBH + HBH & 7.0 & 6.2& 6.2  \\
\hline
IBH + AGN &7.2 & 6.5 & 6.7\\
\hline
IBH + PBH & 6.5 & 5.7 & 5.9  \\
\hline
HBH + PBH & 7.0 & 5.6 & 5.7 \\
\hline
\hline
\multicolumn{4}{|c|}{3 populations}\\
\hline

\texttt{Model~I}: IBH+HBH+PBH & 7.6 & 6.4 & 6.4 \\
\hline
\texttt{Model~II}: IBH+HBH+PBH & 7.4 & 6.4 & 6.4 \\
\hline
\end{tabular}
\caption{Log$_{10}$ Bayes factors (second column), ratio of maximum likelihood (third column), and ratio of average  likelihood (fourth column), for each combination of the models considered in this work, relative to the \texttt{Gaussian Component Spins} spin model for GWTC-4.0, identified as $\bigstar$ as in~\cite{LIGOScientific:2025pvj} (see Eq.~\eqref{ratiosL} for the definitions of ${\cal R}^{\cal L}_{\rm max}$ and ${\cal R}^{\cal L}_{\rm av}$). Results shown are derived including the $\sigma^2_{\ln \hat {\cal L}}$ cut in each Bayesian inference.}
\label{tab:GWTC-4_BFs}
\end{table}

\subsection{Model comparison}
\label{sec:model_comparison}

In Table~\ref{tab:GWTC-4_BFs} we report the estimated $\rm log_{10}$ Bayes factors for each of the scenarios considered. 
The reference model is taken to be the \texttt{Gaussian Component Spins} model, 
denoted by $\bigstar$ and employed in the LVK analysis~\cite{LIGOScientific:2025pvj}, which does not include any correlation between the binary component masses and their spins. The Bayes factor values were computed by taking into account of the effective volume explored by the samples \cite{Mould:2025dts}. Thus, they are defined as:
\begin{equation}
    \mathcal{B}^{\mathcal{M}}_{\bigstar}=\frac{\mathcal{Z}^\mathcal{M}}{\mathcal{Z}^{\bigstar}} \cdot \frac{\mathcal{V}_{\rm eff}^\mathcal{M}}{\mathcal{V}_{\rm eff}^{\bigstar}}
\end{equation}
being $\mathcal{Z}^{\cal M}$ and $\mathcal{V}^{\cal M}_{\rm eff}$ respectively the evidence and the effective volume of a given model ${\cal M}$.

In Table~\ref{tab:GWTC-4_BFs} we also list the ratio between the maximum and posterior-averaged likelihoods found by the model and those of the \texttt{Gaussian Component Spins} model, namely
\begin{align}
    {\cal R}^{\cal L}_{\rm max} \equiv \log_{10} \frac{{\rm max}\ {\cal L}_{\mathcal{M}}}{{\rm max}\ {\cal L}_{\bigstar}}\,, \qquad
    {\cal R}^{\cal L}_{\rm av} \equiv \log_{10} \frac{\langle {\cal L} \rangle_{\mathcal{M}}}{\langle {\cal L} \rangle_{\bigstar}}\,. \label{ratiosL}
\end{align}

We examine the Bayes factors together with the maximum and average likelihoods to assess whether any model is preferred by the GW data.
While Bayes factors are inevitably affected by the choice of priors, reporting information about the likelihood allows us to more robustly interpret how well each model, with a varying number of parameters, improves the inference. As the max-likelihood ratios and the ratios of average likelihoods show similar values to the Bayes factors, we are confident that the effect from the prior volume remains subdominant.

When including only one formation channel, the results show that the IBH, HBH, and AGN channels are decisively preferred with respect to the \texttt{Gaussian Component Spins} model, indicating that the catalog strongly favors models featuring mass-spin correlations, in agreement with the conclusions of Ref.~\cite{Pierra:2024fbl}. 
Among the various models composed of a single population, the AGN channel seems to provide the best fit, with only a slight (but not statistically conclusive) preference over the HBH channel. 
This small advantage of the AGN scenario likely arises from its preferentially aligned spin directions, which are only marginally favored by the data~\cite{LIGOScientific:2025pvj}. 
Both AGN and HBH are favored over the IBH scenario, with a relative difference of respectively $\log_{10}(\mathcal{B}^{\mathrm{HBH}}_{\mathrm{IBH}})=0.9$, and $\log_{10}(\mathcal{B}^{\mathrm{AGN}}_{\mathrm{IBH}})=1.6$. 
No Bayes factor is reported for the PBH-only scenario, since the analysis does not converge due to the likelihood variance cut, which prevents an adequate exploration of the parameter space and thus reliable fits. This is because the PBH channel struggles to reproduce mildly spinning BHs at low masses without simultaneously over-predicting large spins at higher masses. Explaining the low-mass events, in fact, requires efficient accretion (i.e., low $z_\text{cut-off}$), which in turn predicts large spins across the entire mass range. In short, a single population of PBHs cannot describe the observed spins from GWTC-4.0. 

When combining two formation channels, both the IBH+AGN and IBH+HBH models perform comparably to the AGN or HBH channels alone. This already suggests that, in these mixed scenarios, the IBH component is subdominant relative to the dynamical channels. This conclusion is further supported by Bayes factors of $\log_{10}\mathcal{B} = \mathcal{O}(2)$ with respect to the IBH-only case, indicating a strong preference for including at least one dynamical channel.

Interestingly, the data favor a spin magnitude distribution characteristic of hierarchical mergers, whether HBH or AGN, while also showing a mild preference for an aligned-spin component. In the IBH+AGN model this component is supplied by the AGN channel, whereas in the IBH+HBH model it is provided by the IBH subpopulation. 

The 2-population model IBH+PBH, which does not include hierarchical mergers (i.e., either HBH or AGN), is essentially equivalent to the HBH model alone. 
We also explore configurations in which the PBH channel is added to either the IBH or HBH populations. In both cases, the inclusion of a PBH component increases the $\log_{10}$ Bayes factor by approximately one. We do not consider the HBH+AGN combination, as these channels are degenerate apart from their spin-orientation distributions, and thus this scenario is expected to yield evidence comparable to that of the individual channels.

Finally, we consider the 3-channel combination IBH+HBH+PBH, for \texttt{Model~I} and \texttt{Model~II}, respectively. 
For \texttt{Model~I}, which assumes a common redshift distribution for all channels, the inclusion of an additional population is preferred, but with a low statistical significance.
Indeed, the 3-channel model is preferred by merely $\log_{10}\mathcal{B} \sim \mathcal{O}(0.4)$ with respect to the best 2-population case, IBH+AGN. When relaxing the assumption of a common redshift distribution with \texttt{Model~II}, we find only minor differences between that and the 2-population scenario.
This outcome simply reflects the fact that (i) information on the merger-rate redshift evolution remains subdominant compared to spin information in the current catalog, and/or (ii) the data do not require each subpopulation to follow radically different redshift distributions.

\begin{figure*}[!t]
    \centering
\includegraphics[width=.41\linewidth]{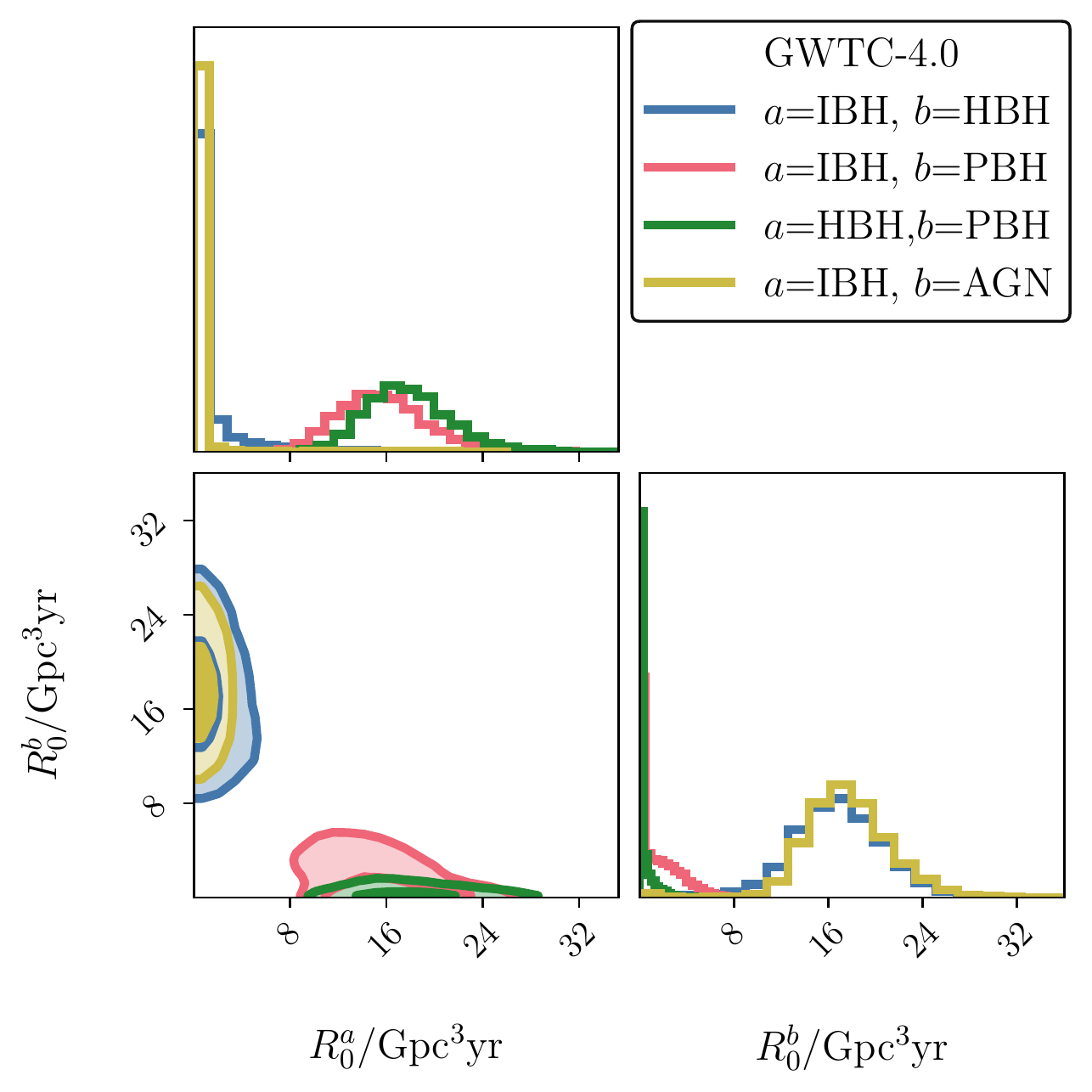}
\includegraphics[width=0.57\linewidth, trim  = 0 -0.04cm 0 0]{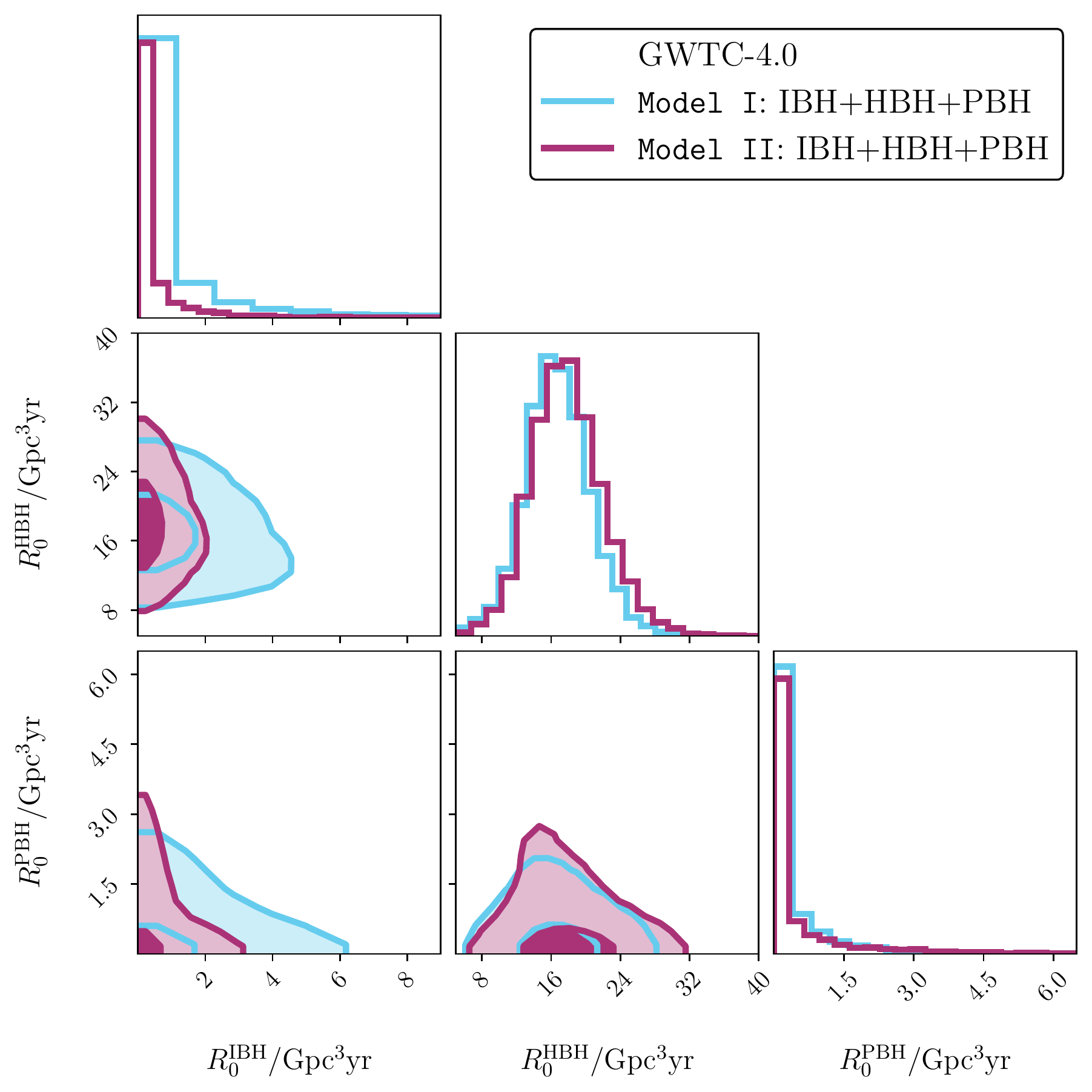}
    \caption{
Left panel: Local merger rate density for the 2-population scenarios considered in this work: IBH+HBH, IBH+AGN, IBH+PBH, and HBH+PBH. The first (second) column corresponds to the first and second subpopulation in each case.
Right panel: Local merger rate density for the 3-population scenarios. 
In both plots, the contours in the 2D-posteriors denote the 68\% and 95\% credible regions. }
\label{fig:GWTC-4_rates}
\end{figure*}

\begin{figure*}[t]
    \centering
\includegraphics[width=1\textwidth]{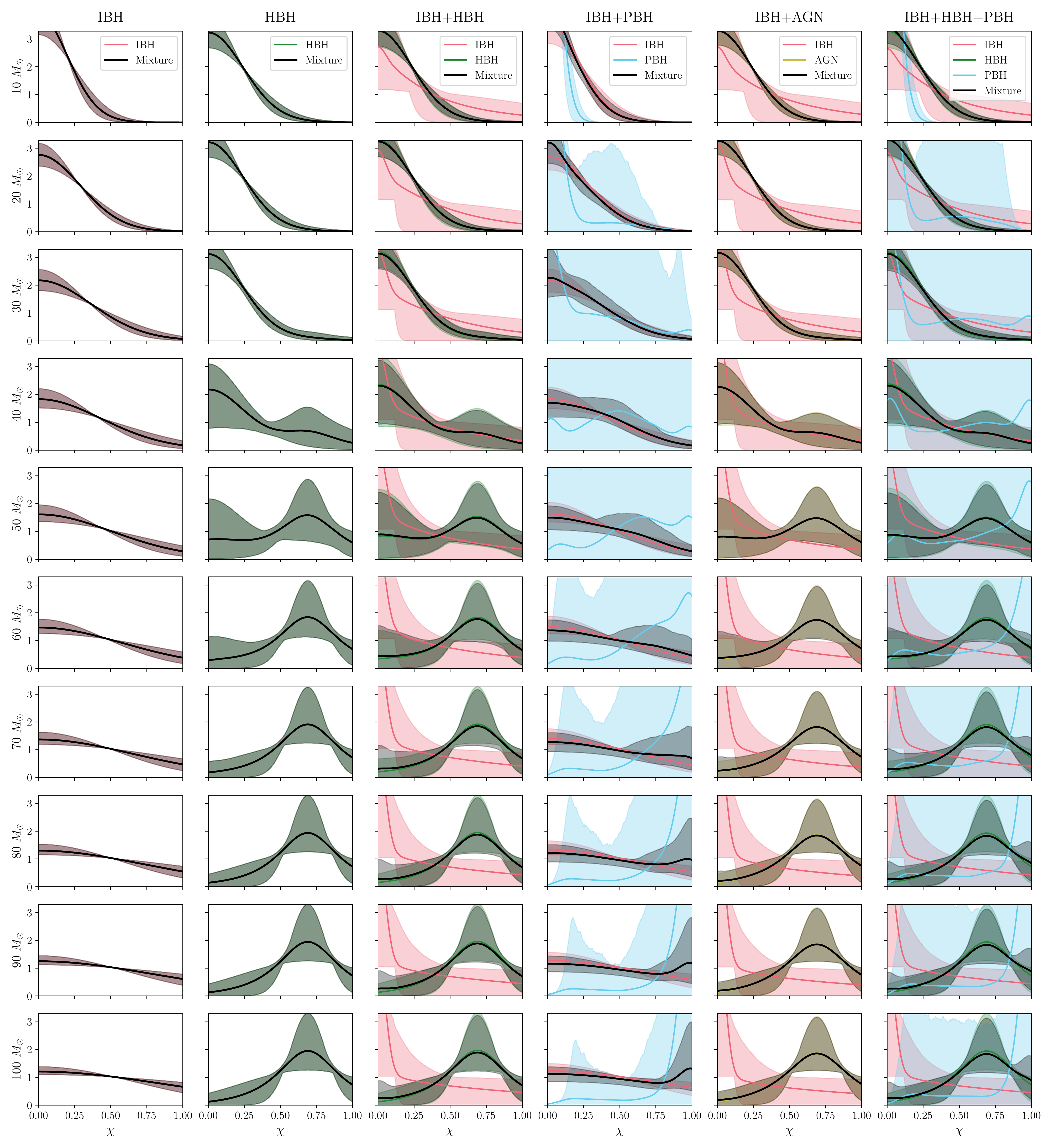}
    \caption{
     Marginalized 1D PPDs of the spin magnitudes $\chi$ for different mass ranges, from $10$ to $100\,M_{\odot}$. Continuous colored lines represent the median value for each subpopulation, while the shaded region shows the $90\%$ credible interval.
     The black line shows the mixture of all channels. Each plot corresponds to \texttt{Model~I}, in which all populations share the same merger-rate redshift evolution. We do not show the AGN case, which yields very similar results to the HBH PPDs, as observed when comparing IBH+HBH with IBH+AGN.
     }
    \label{fig:GWTC-4_posterior_predictive}
\end{figure*}

Considering the 2-population and 3-population models, we are interested in what fraction of each possible formation channels contributes to the overall BBH population.
The corner plots in Fig.~\ref{fig:GWTC-4_rates} display the local merger rate densities for the 2- and 3-population scenarios, respectively. 
Whenever present, both the HBH and AGN channels provide the dominant contribution to the merger rate, with $R_0 \sim 20\,{\rm Gpc}^{-3}\,{\rm yr}^{-1}$, a value similar to (and fully consistent with) the total merger rate density inferred in~\cite{LIGOScientific:2025pvj}. 
Only in the 2-population IBH+PBH case we find that both channels may contribute comparably to the overall merger rate, with the PBH component being subdominant and slightly anti-correlated with the IBH rate.
In the 3-population scenario, we also find that the hierarchical channel is the dominant one, while only an upper limit can be set on the others.
Furthermore, only very mild differences in the mixing fractions are observed between \texttt{Model~I} and \texttt{Model~II}, again showing that the inclusion of individual rate evolution currently provides less constraining power compared to spin information.

We also repeat a subset of our analysis using the GWTC-3 catalog to compare with the results obtained from the latest GWTC-4.0 dataset. While GWTC-3 already showed hints that mass-spin correlations were favored~\cite{Franciolini:2022iaa,Pierra:2024fbl,Santini:2023ukl, Sadiq:2025vly}, the evidence was not strong enough. The larger GWTC-4.0 catalog now allows us to reach more robust conclusions.
Overall, the Bayes factors derived from GWTC-4.0 confirm the trends observed with GWTC-3, once the expected scaling of model preference with the number of detected events is taken into account. 
For example, \texttt{Model~I} exhibited a Bayes factor of about 3 relative to the reference model in GWTC-3, which included 69 events. 
Given the 153 events in GWTC-4.0, one would naively expect the Bayes factor to increase approximately as $3 \times 153/69 \simeq 6.7$, assuming the two catalogs have comparable statistical properties and each event is statistically independent. 
This expectation is indeed consistent with the value of $\sim 7.5$ reported in Table~\ref{tab:GWTC-4_BFs} for a fixed set of priors. 

To summarize, our findings are the following: (i) models with mass-spin correlation are strongly preferred with respect to the uncorrelated models~\cite{LIGOScientific:2025pvj}; 
(ii) when compared to single channels, the ones with multiple populations are only slightly preferred, or comparable to the 1-population HBH/AGN cases; 
3) adding a third channel only mildly improves the fit; 
and 4) adding different redshift-evolution distributions for the various channels has negligible impact, suggesting that redshift information is sub-leading relative to spin information in the GWTC-4.0 catalog.

\subsection{Posterior predictive distributions}
\label{ppd}

In this section, we examine the posterior predictive distributions (PPDs), which represent the reconstructed astrophysical distributions of observables---here, the spin magnitudes and tilt angles---after accounting for selection effects. Those distributions are predicted by a given model after accounting for the uncertainties in its inferred parameters, and provide a direct way to visualize possible differences between competing scenarios.

\subsubsection{Spin parameters}

In Fig.~\ref{fig:GWTC-4_posterior_predictive}, we show the PPD for the spin magnitude $\chi$ (x-axis) as a function of the BH mass (different panels on the y-axis), ranging from $10\,M_\odot$ to $100\,M_\odot$, and for different combinations of populations. This allows us to visualize how the spin distribution evolves across the mass spectrum. From left to right, each column corresponds to a different population scenario, ranging from a single population to three populations, all within the \texttt{Model~I} framework. A general trend visible in most scenarios is that low-mass BHs tend to have low spin magnitudes, with the distribution shifting toward higher spins at higher masses, in agreement with~\cite{Pierra:2024fbl,Li:2023yyt}.

We start with the single-population results. 
For IBHs, the spin distribution is sharply peaked at small spin values at low masses, and gradually flattens toward higher $\chi$ at larger masses. 
In contrast, the HBH model starts showing a distinct secondary peak at $\chi \simeq 0.7$ around $40\,M_\odot$, associated with the 2g component (see Eq.~\ref{p_chi_1g2g}), which becomes dominant above $\sim 60\,M_\odot$. 
Further, the AGN model gives similar results to HBH in terms of spin magnitude, as the different assumptions on spin orientation do not affect the inferred spin amplitude. For this reason we do not show the AGN PPD.
We note that the appearance of a peak at $\chi = 0.7$ is imposed by construction in our HBH/AGN models (as for 2g the mean of the Gaussian is fixed at $\sim 0.7$). However, the model also includes a width for this peak, with a prior large enough to allow for this distribution to potentially become effectively uniform. Therefore, the preference for a relatively sharp 2g peak is driven by the data. 
The higher Bayes factors obtained for HBH and AGN arise from their ability to predict larger spins above $\sim 50\,M_\odot$. 

Except for the IBH case, which shows only a mild dependence on the mass due to its Gaussian-like spin magnitude distribution, a clear transition is observed around $m = \mathcal{O}(40\text{--}50)\,M_\odot$ for all the models. 
At small masses, $\chi$ tends to be low and peaks near zero, while at intermediate masses a secondary peak develops at larger spins, leading to a distribution that definitely prefers larger spins in the heaviest mass range, as found in~\cite{Pierra:2024fbl}.

We now consider the 2-population cases: IBH+HBH, IBH+PBH, and IBH+AGN. 
In the combinations that include IBH with either HBH or AGN, those channels dominate, and the resulting spin distributions closely resemble those of HBH or AGN alone. 
This confirms what was found in the previous sections based on the Bayes factors and local merger rates, namely that the HBH (or AGN) channel provides the main contribution. 
In both the IBH+HBH and IBH+AGN scenarios, the IBH population rate is largely subdominant, leaving the corresponding population parameters essentially unconstrained. In practice, the HBH/AGN channel accounts for nearly all the events, while the IBH component contributes only a negligible fraction, preventing any meaningful constraint on its parameters. The shaded band corresponds to the 90\% credible interval obtained by randomly sampling the Gaussian spin distribution model \eqref{eq:spin_IBH} with variance within the prior range 0.1 and 1, reflecting the width allowed by our prior. 
In the IBH+PBH case, both populations contribute comparably, with PBHs reproducing the enhanced spin probability at masses above $\sim 40\,M_\odot$. 
A characteristic feature of the PBH model is that, after the transition from small masses and small spins to larger masses and higher spins, the heaviest mergers are predicted to have very large spins, with the distribution developing a peak close to $\chi \simeq 1$. 
While this feature partly results from the extrapolation of the model, it is particularly compatible by the high-spin event GW231123~\cite{LIGOScientific:2025rsn}. 
However, as we verified by repeating the analysis without including this event in the catalog, the inference in this scenario is not dominated by the inclusion of GW231123 (see also Appendix~\ref{app:masspos}). 

\begin{figure*}[t]
    \centering
\includegraphics[width=1\textwidth]{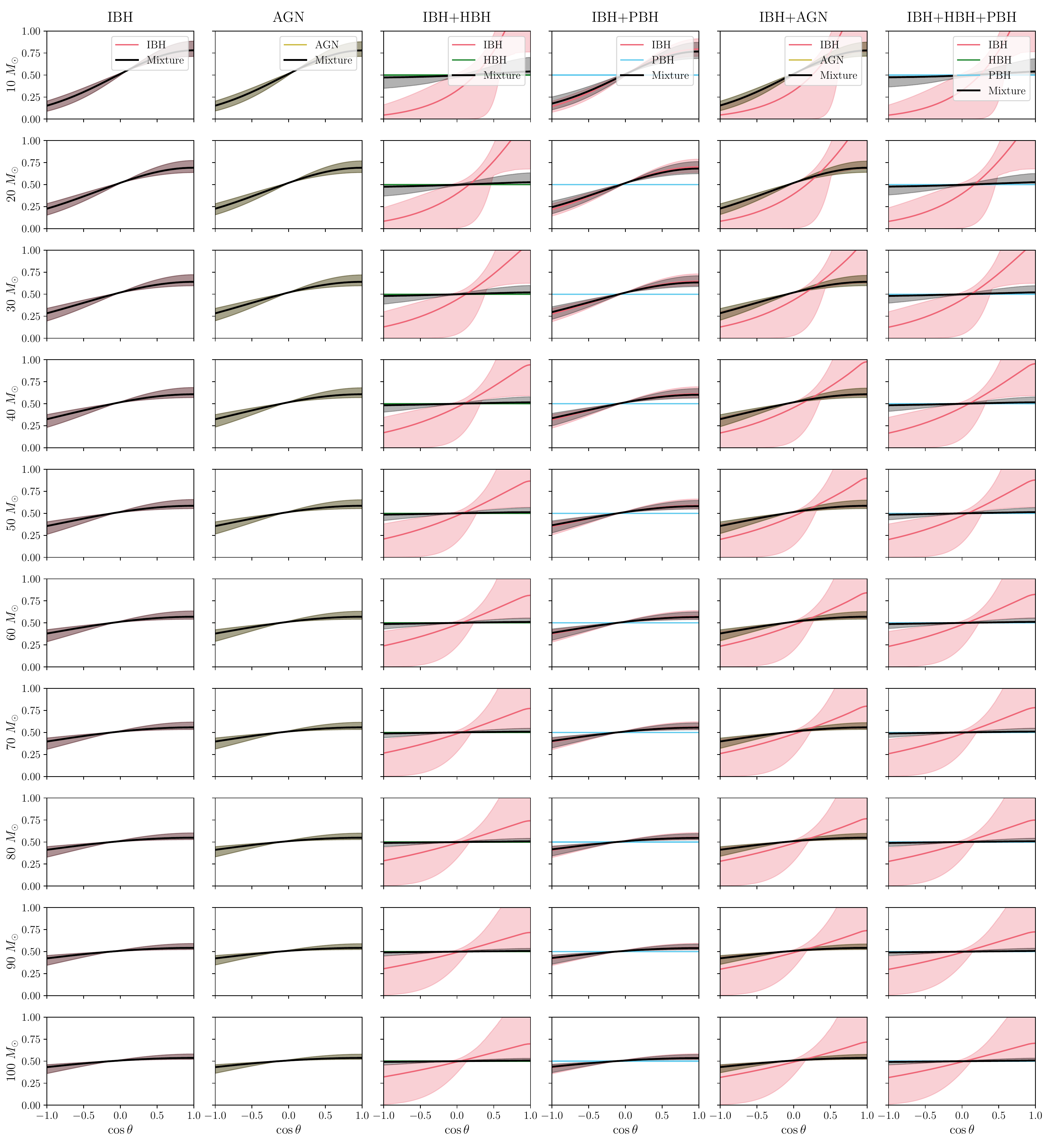}
    \caption{
Marginalized 1D PPDs of the polar angle $\cos \theta$ for different mass ranges, from $10$ to $100\,M_{\odot}$. The color notation is the same used in 
    Fig.~\ref{fig:GWTC-4_posterior_predictive}.
    In the second column, we show the AGN model instead of the HBH model, as the latter force the distribution to be $p(\cos \theta) = 1/2$.
    }
    \label{fig:GWTC-4_posterior_predictive_costheta}
\end{figure*}

\begin{figure*}[t]
    \centering
\includegraphics[width=0.49\textwidth]{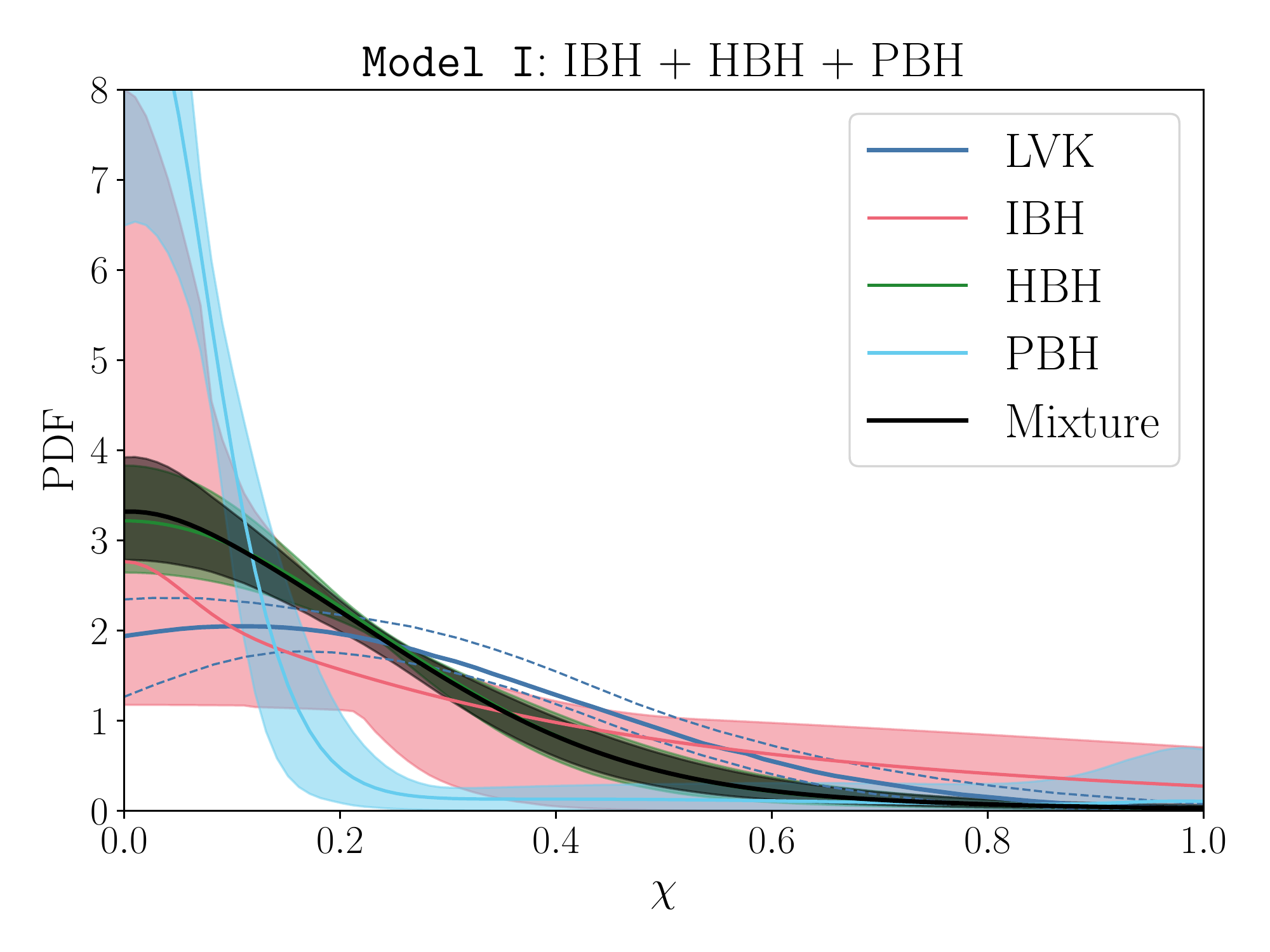}
\includegraphics[width=0.49\textwidth]{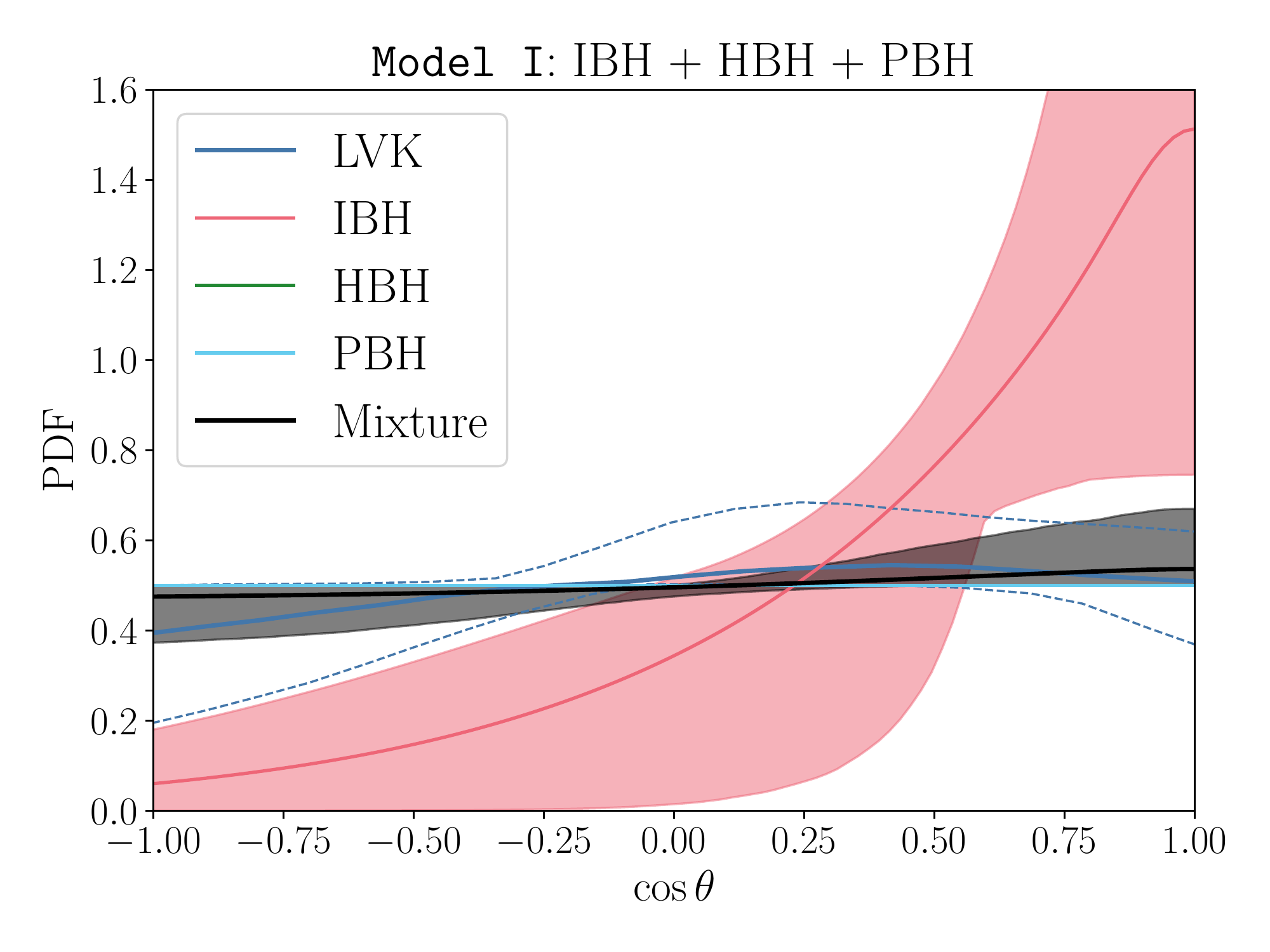}
    \caption{Marginalized 1D PPD of the spin magnitudes $\chi$ (left) and the cosine of the tilt angles (right), for \texttt{Model~I} and averaged over the mass distribution. Continuous lines show the median value of the PPD, and the shaded regions represent the 90\% C.L. region. We show the IBH+HBH+PBH combination in black, and the individual contribution of each channel to the 3-population fit using different color schemes. 
    The blue dashed line is the result obtained with the \texttt{Gaussian Spin Model} adopted by the LVK Collaboration in~\cite{LIGOScientific:2025pvj}.}
    \label{fig:GWTC-4_chi_t_mm}
\end{figure*}

Finally, for the 3-population scenarios, we report only the results from \texttt{Model~I}, since those obtained with \texttt{Model~II} are very similar.
In the 3-population case, we again find the dynamical channel to be dominant, with only a small contribution from the IBH and PBH populations, the latter being especially relevant in reproducing the peak at very large spins and high masses favored by the inclusion of the very massive event GW231123.

Similar qualitative trends are observed in Fig.~\ref{fig:GWTC-4_posterior_predictive_costheta}, although the interpretation of the $\cos\theta$ distributions requires some care. The figure indicates that the tilt angle appears to be rather well constrained and that, for most models, it shows little evolution with mass. This behavior depends on the the adopted parametrizations, especially for the PBH and HBH channels, whose tilted-spin distributions is flat and mass-independent.

For the IBH model, the PPD suggests a preference for positive tilts at low masses, while at higher masses it gradually approaches the flat distributions predicted by dynamical channels (HBH) and by the PBH population. Since the HBH and PBH models predict flat tilt-angle distributions, only IBH/AGN can accommodate such a feature. 
Overall, the main conclusions that can be drawn from this figure are the following:  
(i) the choice of the model parametrizations adopted tends to constrain the tilt-angle distribution quite strongly;  
(ii) there is no compelling evidence for a mass-dependent evolution of $\cos\theta$, except for a possible mild trend in the IBH model, whose significance remains uncertain;  
(iii) current data do not allow us to distinguish between the specific assumptions for spin-orientation distributions, confirming that the information contained in $\chi$ remains the main driver of model differentiation.


In Fig.~\ref{fig:GWTC-4_chi_t_mm}, we show the PPD for both $\chi$ and $\cos \theta$, marginalized over the mass distribution.
Since the overall mass distribution is more populated at low masses, the resulting averaged distributions largely resemble that of low-mass events shown previously in Figs.~\ref{fig:GWTC-4_posterior_predictive} and \ref{fig:GWTC-4_posterior_predictive_costheta}.
However, this marginalized plot does not capture the mass dependence present in the model.  
While in Fig.~\ref{fig:GWTC-4_chi_t_mm} we only show the best fitting scenario \texttt{Model~I}: IBH+HBH+PBH,
the plots obtained in the other cases are similar, i.e. they show that the marginalized distributions are very similar to the low-mass PPD shown in Figs.~\ref{fig:GWTC-4_posterior_predictive} and \ref{fig:GWTC-4_posterior_predictive_costheta}.

The blue lines indicate the mean and $90\%$ C.L. obtained using the \texttt{Gaussian Spin Model} in~\cite{LIGOScientific:2025pvj}. In Appendix~\ref{app:LVKmodel} we give more details on this model, and show how we reproduced the LVK results.
Looking first at the spin magnitudes, there is one visible difference with the \texttt{Gaussian Spin Model} model: it seems to be preferring slightly larger values of the spin magnitude.
This is likely due to the absence of mass-spin correlations in the \texttt{Gaussian Spin Model} model, which is compensated by attributing the preference for spinning events---typically observed at higher masses---to an overall shift toward larger spin magnitudes. 

Regarding the marginalized distribution of the tilt angle $\cos\theta$, the IBH model does not allow the peak of the distribution to deviate from $\cos\theta = 1$. However, the combined model is constructed as a mixture of IBH and HBH/PBH components, with the latter predicting a flat distribution in $\cos\theta$. In the three-population scenario, the IBH contribution is allowed to be negligible by the rate priors, but appears to play a role. The small excess toward positive alignment that we observe at low masses naturally propagates into the marginalized distribution. Nonetheless, the result remains fully compatible, within the 90\% credible interval, with a flat distribution ($1/2$ at $\cos\theta = 1$). Thus, we do not find evidence for alignment, only a mild trend.
Moreover, the inferred distribution is consistent with the LVK result within uncertainties, indicating a much weaker tension than what is observed for $\chi$. This highlights that spin-orientation measurements are only weakly informative: different prior assumptions and model choices can describe the data equally well, and models without mass-orientation correlations (such as the one adopted by the LVK Collaboration) remain compatible with current observations.

\begin{figure*}[t]
    \centering
\includegraphics[width=0.49\textwidth]{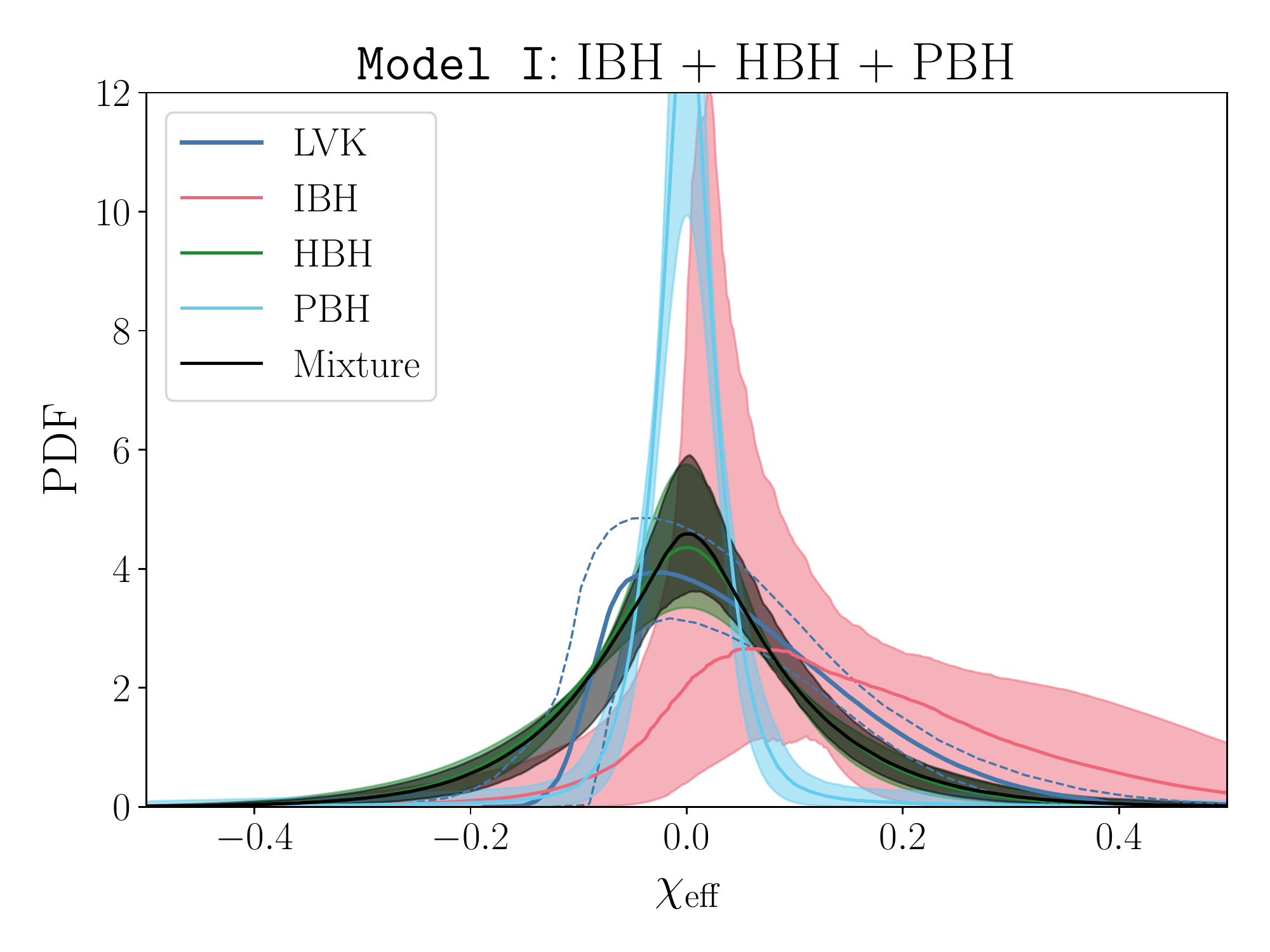}
\includegraphics[width=0.49\textwidth]{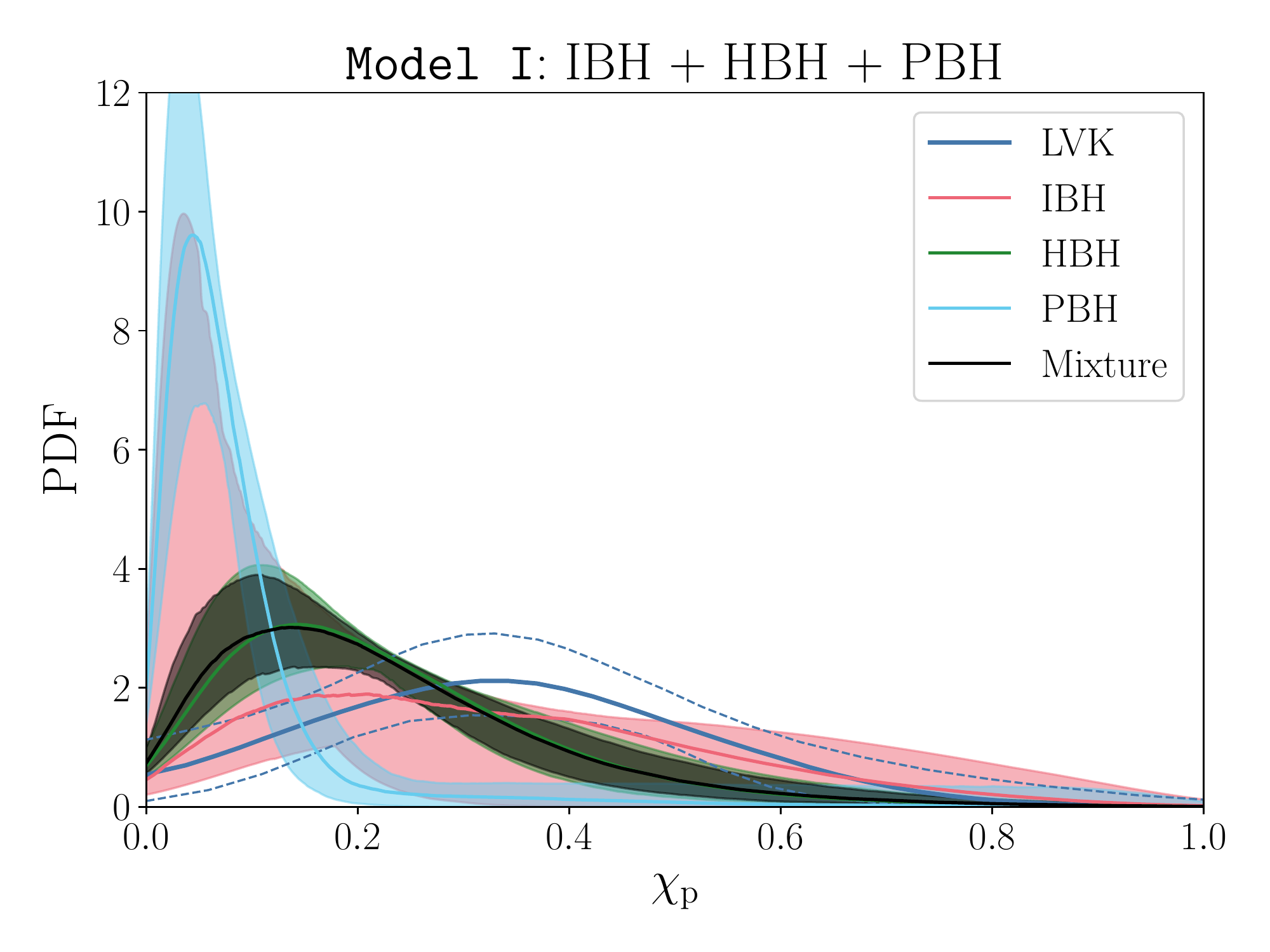}
    \caption{ 
    Marginalized 1D PPD for the effective spin parameter $\chi_{\rm eff}$ and the effective precession parameter $\chi_{\rm p}$.
    The color notation is the same introduced in Fig.~\ref{fig:GWTC-4_chi_t_mm}.}
    \label{fig:GWTC-4_posterior_predictive_chieff_chip}
\end{figure*}

Finally, in Fig.~\ref{fig:GWTC-4_posterior_predictive_chieff_chip}, we present the PPDs of the effective spin $\chi_{\mathrm{eff}}$ and of the spin-precession parameter $\chi_{\rm p}$, under the same assumptions as in Fig.~\ref{fig:GWTC-4_chi_t_mm}. Like before, the distributions are obtained by averaging over the intrinsic mass distribution of the BBH population, and are therefore dominated by low-mass events. While this plot contains the same physical information as the spin-magnitude and tilt-angle distributions, it is useful to compare it directly with the phenomenological fits provided by the LVK Collaboration. In particular, for $\chi_{\mathrm{eff}}$, the LVK \texttt{Skew-normal Effective Spin} fit (blue) predicts an asymmetric distribution slightly shifted toward larger spin magnitudes compared to the mixture model, whereas the HBH and PBH contributions in our framework keep the PPD more symmetrically distributed around zero. The only asymmetric contribution is introduced by the IBH model, which skews the distribution towards positive values (see also the discussion in~\cite{Banagiri:2025dxo}). 
For $\chi_{\rm p}$, we find that the distribution peaks around $\chi_{\rm p} \sim 0.1$, with some mild tension with the LVK PPD, which prefers larger values of $\chi_{\rm p}$. This is probably explained by the fact that our models prefer smaller spin magnitude for the light events that dominate the intrinsic population (that does not take into account selection effects).

\section{Conclusions}\label{sec:conclusions}

In this work, we have investigated the presence of physically motivated correlations between BH masses and spin magnitudes in BBH merger events from the GWTC-4.0 catalog recently released by the LVK collaboration. 

Compared to GWTC-3, we found that GWTC-4.0 shows stronger evidence for the presence of a mass-spin magnitude correlation, a conclusion which is consistent with the expected statistical improvement predicted with the increased number of GW detections.
We showed that, among the models we considered, only the dynamical channels (HBH or AGN) provide a competitive fit when treated as single-population scenarios. Multi-population models can slightly improve the fit, allowing the remaining channels to contribute. However, once included, HBH and AGN consistently dominate the inference, primarily due to the preferred shape of the individual-spin distribution as a function of mass.
In those cases, the other channels, such as IBH or PBH, are constrained to contribute only marginally, typically through upper limits.
If present, the PBH channel predicts a subpopulation of near-extremal BHs at high masses. This is consistent with events such as GW231123, although the overall fit is not driven by this single event.

From the reconstructed spin distributions, we observed a transition between spin distributions around $M \simeq 40$--$50\,M_\odot$. 
This pattern could be explained by the hierarchical mergers (i.e., the HBH or AGN channels), while the IBH population dominates only at small masses and low spins. 
The PBH population, when present, can also reproduce the high-spin regime at large masses. 
For the spin orientations, only the IBH/AGN models can explain the preference for positive tilts at small masses, while at high masses the PPDs converge to those predicted by the HBH and PBH channels. No strong preference between the HBH and AGN scenarios is found, indicating that current data are more sensitive to spin magnitudes than to spin orientations. This reduced sensitivity to spin orientation is consistent with the fact that this parameter is only weakly constrained by the GW signal itself.

In this work, we focused on events from the GWTC-4.0 catalog and neglected the recent highly spinning events GW241110 and GW241011, with masses below 30 $M_\odot$~\cite{LIGOScientific:2025brd}. With the inclusions of such events in the catalog, we could expect the support for dynamical channels to grow. However, we note that despite their peculiar high spins, there is no evidence for these events to be population outliers, in agreement with Ref.~\cite{LIGOScientific:2025brd}. 
In this paper we have focused on monotonic mass transitions between spin magnitude distributions. A recent analysis of GWTC-4.0, focusing on the effective spin parameter $\chi_{\rm eff}$, argued for the presence of two sub-populations of 2g BHs (possibly with higher and more isotropic spins) around $\sim 14 M_\odot$ and above $45 M_\odot$~\cite{Tong:2025xir}. 
Another work reported evidence for three subpopulations in the GWTC-4.0 catalog---i.e., two populations with small spin amplitude at masses below $\sim 40 M_\odot$ (with different mass ratio distributions), and one with a flat spin distribution at higher masses~\cite{Banagiri:2025dmy}. Both studies are consistent with our findings for a spin transition around $40 M_\odot$. The extension of population models to include other possible sub-populations is left for future work.

Finally, it is worth stressing that our approach to constrain the mixing fractions between models was overly conservative, since it is based on discriminating between models using only with mass-spin correlations. While we have shown that including different merger rate evolutions for different models has a minor impact on the analysis, the different mass distributions associated with each model do play a major role. 
The investigation of more general models, in which the assumption of a common mass distribution is also relaxed, is an important topic for future work.

\begin{acknowledgments}
We thank Alessandro Agapito, Sarah Ferraiuolo, Stephen Green, Konstantinos Kritos, Francesco Iacovelli, Leonardo Iampieri, Matthew Mould, and Luca Reali for discussions. 
E.B. is supported by NSF Grants No.~AST-2307146, No.~PHY-2513337, No.~PHY-090003, and No.~PHY-20043, by NASA Grant No.~21-ATP21-0010, by John Templeton Foundation Grant No.~62840, by the Simons Foundation [MPS-SIP-00001698, E.B.], by the Simons Foundation International [SFI-MPS-BH-00012593-02], and by Italian Ministry of Foreign Affairs and International Cooperation Grant No.~PGR01167.
This work was carried out at the Advanced Research Computing at Hopkins (ARCH) core facility (\url{https://www.arch.jhu.edu/}), which is supported by the NSF Grant No.~OAC-1920103.
F.C. acknowledges the financial support provided under the ``Progetti per Avvio alla Ricerca Tipo 1,'' protocol number AR12419073C0A82B. 
G.F. thanks IFPU and the organizers of the workshop \textit{"Primordial BHs in the Multi-Messenger Era"} for the stimulating environment where part of this work was carried out and first presented.
S.M. and G.P. are supported by the ERC grant GravitySirens  101163912. Funded by the European Union. Views and opinions expressed are however those of the author(s) only and do not necessarily reflect those of the European Union or the European Research Council Executive Agency. Neither the European Union nor the granting authority can be held responsible for them.
P.P. is supported by the MUR FIS2 Advanced Grant ET-NOW (CUP:~B53C25001080001) and by the INFN TEONGRAV initiative.
Some numerical computations were performed at the Vera cluster, supported by MUR and Sapienza University of Rome.
This material is based upon work supported by NSF's LIGO Laboratory which is a major facility fully funded by the National Science Foundation.
\end{acknowledgments}

\appendix

\section{The LVK Gaussian component spin model} \label{app:LVKmodel}

In this section, we present the reference spin models adopted in the population analysis 
carried out by the LVK Collaboration in the GWTC-4.0 release~\cite{LIGOScientific:2025pvj}.

Following the most recent analysis by the LVK Collaboration, we model the spin magnitudes $\chi_i$ as a truncated Gaussian distribution between 0 and 1, assuming they are identically and independently distributed. Therefore, 
\begin{equation}
p(\chi_1, \chi_2 | \mu_\chi, \sigma_\chi) = \mathcal{N}_{[0,1]}(\chi_1 | \mu_\chi, \sigma_\chi) \mathcal{N}_{[0,1]}(\chi_2 | \mu_\chi, \sigma_\chi).
\end{equation}

\begin{figure}[t]
    \centering
\includegraphics[width = 0.49 \textwidth]{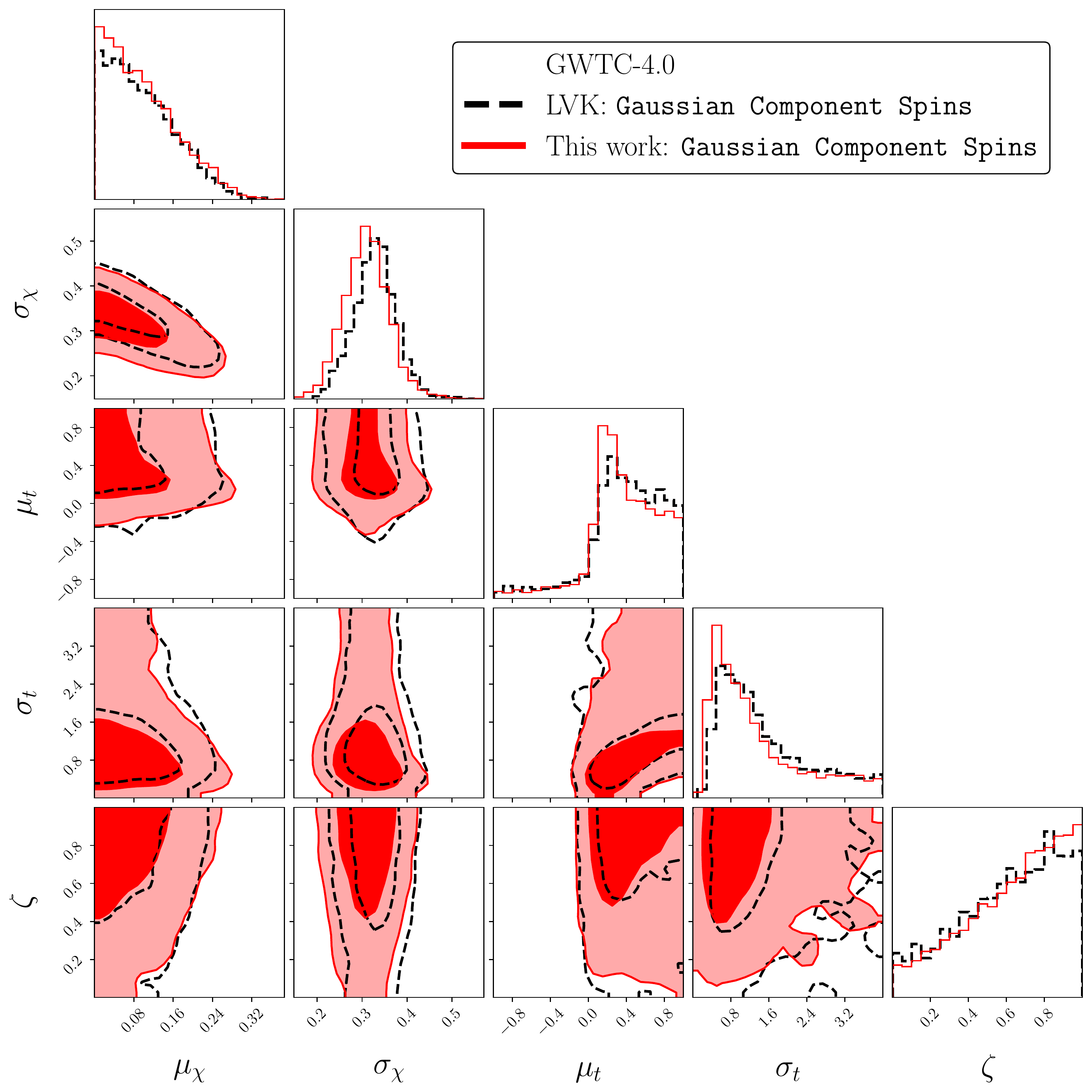}
    \caption{ 
    Posterior distribution for the reference spin models considered in the GWTC-4.0 LVK population analysis. In red, we report the posterior distribution we obtained, while black dashed lines show the LVK posteriors~\cite{ligo_scientific_collaboration_2025_16911563}.
    }
    \label{fig:default_spin_models LVK_pos}
\end{figure}

We model the spin inclination distribution as a mixture between a Gaussian distribution truncated on $[-1, 1]$ and an isotropic distribution, assuming they are identically but \emph{not} independently distributed~\cite{KAGRA:2021duu}: 
\begin{align}
p(\cos \theta_1,\cos\theta_2) 
&= (1 - \zeta) \times  \left (\frac{1}{4} \right )
\nonumber \\
&+ \zeta \mathcal{N}_{[-1,1]}(\cos \theta_1 | \mu_t, \sigma_t) \mathcal{N}_{[-1,1]}(\cos \theta_2 | \mu_t, \sigma_t) 
.
\end{align}
We report the priors on the Gaussian component spins we adopt in Table~\ref{tab:priors_spin_Gaussian}, and the corner plot of the hyperparameters in Fig. \ref{fig:default_spin_models LVK_pos}.

\begin{table}[h!]
\centering
\caption{Priors on the BBH spin and redshift-evolution hyperparameters for the \texttt{Gaussian component spin model}.}
\label{tab:priors_spin_Gaussian}
\begin{tabularx}{\linewidth}{|X|l|l|}
\hline\hline
Spin model & Parameter & Prior \\
\hline\hline
\multirow{5}{*}{Default spin model}
  & $\mu_\chi$ & ${\cal U}[0,\,1]$ \\
  & $\sigma_\chi$ & ${\cal U}[0.005,\,1]$ \\
  & $\mu_t$ & ${\cal U}[-1,\,1]$ \\
  & $\sigma_t$ & ${\cal U}[0.001,\,4]$ \\
  & $\zeta$ & ${\cal U}[0,\,1]$ \\
\hline\hline
\end{tabularx}
\end{table}

\section{Impact of effective likelihood variance cuts}\label{app:likelihood-cuts}

The hierarchical likelihood is sampled numerically for each population model by combining parameter estimation samples from 
$N_{\text{obs}}$ GW events with a set of detectable injections used to account for selection effects. Both the injections and the parameter estimation samples provide values of the parameters $(\lambda, z)$, which are then employed to evaluate the BBH merger rate. In addition, the priors $\pi_{\text{PE}}$ and $\pi_{\text{inj}}$ 
used to generate the parameter estimation samples and the injections, respectively, must be deconvolved to recover the underlying population properties. The overall log-likelihood can be approximated as
\begin{equation}
\ln \mathcal{L}(\{x\}|\Lambda) \approx 
- \frac{T_{\text{obs}}}{N_{\text{gen}}} 
\sum_{j=1}^{N_{\text{obs}}} s_j
+ \sum_i \ln \left[ 
\frac{T_{\text{obs}}}{N_{s,i}} 
\sum_{j=1}^{N_{s,i}} w_{i,j} \right],
\label{eq:like}
\end{equation}
where $s_j$ and $w_{i,j}$ are the weights associated with the injections 
and the parameter estimation samples, respectively. 
Here, the index $i$ refers to the $i$-th GW event, 
while $j$ labels the Monte Carlo (MC) samples. 
The weights are given by
\begin{align}
s_j &= 
\frac{1}{\pi_{\text{inj}}(\lambda_j)} 
\left. \frac{dN_{\rm BBH}(\Lambda) }{dt \, dz \, d\lambda}\right|_j ,
\label{eq:sj} \\
w_{i,j} &= 
\frac{1}{\pi_{\text{PE}}(\lambda_{i,j}|\Lambda)} 
\left. \frac{d N_{\rm BBH}(\Lambda) }{dt \, d\lambda \, dz} \right|_{i,j}.
\label{eq:wij}
\end{align}

As we evaluate the MC integrals on a finite number of samples from each event and a finite $N_{\mathrm{draw}}$, 
we must carefully account for the intrinsic variance in the likelihood estimation (see e.g.~\cite{Essick:2022ojx}). To assess the reliability of our MC estimators for the likelihood, we follow the recent literature and evaluate the variance of the log-likelihood estimator, which varies across parameter space because of the resampling procedures used in Eqs.~\eqref{eq:sj} and~\eqref{eq:wij}. By propagating this uncertainty along independent degrees of freedom, the variance of the log-likelihood estimator, $\sigma^2_{\ln\hat{\mathcal{L}}}$, for the combined likelihood can be estimated as~\cite{Essick:2022ojx}
\begin{equation}
\sigma^2_{\ln \hat{\mathcal{L}}}(\Lambda) = 
\sum_{i=1}^{N_{\mathrm{det}}} 
\frac{\sigma^2_{\hat{\mathcal{L}}_i}(\Lambda)}{\hat{\mathcal{L}}^2_i(\Lambda)} 
+ N_{\mathrm{det}}^2 \sigma^2_{\xi}(\Lambda),
\end{equation}
where
\begin{equation}
\sigma^2_{\hat{\mathcal{L}}_i}(\Lambda) = \frac{1}{N_{\mathrm{PE}}}
\left[
\frac{1}{N_{\mathrm{PE}} - 1}
\sum_{j=1}^{N_{\mathrm{PE}}}
w_{i,j}^2
- \hat{\mathcal{L}}^2_i(\Lambda)
\right]
\end{equation}
is the MC variance in the single-event integrals of~\eqref{eq:wij}, and 
\begin{equation}
\sigma^2_{\xi}(\Lambda) = \frac{1}{N_{\mathrm{draw}}}
\left[
\frac{1}{N_{\mathrm{draw}} - 1}
\sum_{j=1}^{N_{\mathrm{found}}}
s_j^2
- \xi(\Lambda)^2
\right]
\end{equation}
is the variance in the detection efficiency MC integrals~\eqref{eq:sj}. 
In the previous equations, we introduced $N_{\rm PE}$, $N_{\rm draw}$, and $N_{\rm found}$, denoting respectively the number of posterior samples available for each detected event, the number of population draws used to estimate selection effects, and the subset of those draws that satisfy the detection criteria.

As discussed in Appendix~D3 of~\cite{LIGOScientific:2025pvj} and Appendix~A of~\cite{Mancarella:2025uat}, the inference can be sensitive to the likelihood cuts introduced to ensure reliable likelihood evaluations.
We therefore also adopt a second cut based on the {\it effective number of samples}. We introduce the effective number of posterior samples per event, defined as~\cite{Talbot:2023pex}
\begin{equation}
    N_{\text{eff},i} = \frac{\left( \sum_{j}^{N_{s,i}} w_{i,j} \right)^2}{\sum_{j}^{N_{s,i}} w_{i,j}^2},
\end{equation}
which quantifies how many samples per event are contributing to the evaluation of the integral. In our case, we require to have at least an effective number of posterior samples equal to 20 for each event and population model supported by the analysis. In case this requirement is not satisfied, \texttt{icarogw} will artificially
associate a null likelihood to the specific point in the parameter space of the population model, as it cannot be trusted.
Also, following~\cite{Farr:2019rap}, we impose numerical stability on the injection by defining the quantity
\begin{equation}
    N_{\text{eff,inj}} = \frac{\left[ \sum_{j}^{N_{\text{det}}} s_j \right]^2}{\left[ \sum_{j}^{N_{\text{det}}} s_j^2 - N_{\text{gen}}^{-1} \left( \sum_{j}^{N_{\text{det}}} s_j \right)^2 \right]}.
    \label{eq:neff_inj}
\end{equation}
We impose that $N_{\text{eff,inj}}>4N_{\rm obs}$. However, this cut has been pointed out to be less stringent~\cite{Heinzel:2025ogf}.

In Table~\ref{tab:GWTC-4_BFs_cuts} we compare the Bayes factors obtained using the $\sigma^2_{\ln \hat{\mathcal{L}}}(\Lambda)$ and $N_{\rm eff}$ cuts. The values for the former correspond to those already shown in Table~\ref{tab:GWTC-4_BFs}, and are reported again here side by side for convenience. Overall, the trends remain similar, although the Bayes factors are slightly larger in some cases. The analysis involving the PBH-only model converges and again confirms that this model alone is ruled out as an explanation of the data. 

We also observe that, particularly for models featuring preferentially aligned spins---such as IBH, AGN, IBH+HBH, IBH+AGN, and IBH+PBH---the Bayes factors exhibit a substantial increase. While this result cannot be fully trusted, one may speculate that part of this behavior arises because the variance cut can downweight regions of parameter space associated with tight spin alignment. This effect may also be partly linked to the use of a uniform and isotropic spin prior in the GWTC-4.0 parameter estimation, which yields fewer posterior samples near the strongly aligned configuration (see~\cite{LIGOScientific:2025pvj} for further discussion).

\begin{table}[t!]
\centering
\begin{tabular}{|l|c|c|}
\hline
\textbf{Model $\mathcal{M}$} 
&
\makecell{$\log_{10}(\mathcal{B^{\mathcal{M}}_{\bigstar}})$
\\ $N_{\rm eff}$ cut}
&
\makecell{$\log_{10}(\mathcal{B}^{\mathcal{M}}_{\bigstar})$ 
\\
$\sigma^2_{\ln \hat {\cal L}}$ cut
}\\
\hline
\hline
\multicolumn{3}{|c|}{1 population}\\
\hline
IBH & 5.9 & 5.3 \\
\hline
HBH & 6.8 & 6.2 \\
\hline
AGN & 8.0 & 7.0 \\
\hline
PBH & -12 & -- \\
\hline\hline
\multicolumn{3}{|c|}{2 populations}\\
\hline
IBH + HBH & 7.6 & 7.0 \\
\hline
IBH + AGN & 8.6 &  7.2 \\
\hline
IBH + PBH & 8.5 & 6.5 \\
\hline
HBH + PBH & 7.6 & 7.0 \\
\hline\hline
\multicolumn{3}{|c|}{3 populations}\\
\hline
\texttt{Model~I}: IBH+HBH+PBH & 8.6 & 7.6  \\
\hline
\texttt{Model~II}: IBH+HBH+PBH& 8.7 & 7.4 \\
\hline
\end{tabular}
\caption{Log$_{10}$ Bayes factors relative to the \texttt{Gaussian Component Spins} spin model for GWTC-4.0, identified as $\bigstar$ as in~\cite{LIGOScientific:2025pvj}.
The right column shows the same values reported in Table~\ref{tab:GWTC-4_BFs}, to be compared with the results obtained using the $N_{\rm eff}$ cuts.}
\label{tab:GWTC-4_BFs_cuts}
\end{table}

\section{Mass distribution parameters}\label{app:masspos}

Table~\ref{tab:priors_masses} summarizes the priors adopted for the parameters governing the primary and secondary BH mass distributions adopted for all models in this work. 
These include the slopes of the power-law components, the location and width of the Gaussian peaks, and the lower and upper mass cutoffs. 

The distribution of masses can be further factorized as 
$p_{\rm pop} \equiv p(m_1|\Lambda_m) 
p( m_2|m_1, \Lambda_m)$, where $\Lambda_m$ are the parameters of the mass population. The primary mass $m_{1}$ follows a Power Law + 2 Peaks distribution~\cite{Talbot:2018cva, LIGOScientific:2020kqk}:
\begin{equation}
    \begin{aligned}
p\left(m_{1}| \Lambda_m \right) 
& =(1-\lambda_{g,\text {low }}) \mathcal{P}\left(m_{1} | m_{\min }, m_{\max },-\alpha\right)
\\
& + \lambda_g \lambda_{g,\text {low }} \mathcal{G}\left(m_{1} | \mu_{g}^{\text {low }}, \sigma_{g}^{\text {low }}\right)
\\
& +\lambda_g\left(1-\lambda_{g,\text {low }}\right) \mathcal{G}\left(m_{1} | \mu_{g}^{\text {high }}, \sigma_{g}^{\text {high }}\right),
\end{aligned}
\end{equation}
where $\mathcal{P}$ is a power law truncated between $m_{\min}$ and $m_{\max}$ with slope $\alpha$, and $\mathcal{G}$ is a Gaussian distribution centered at $\mu_g$ with standard deviation $\sigma$. The mixing fractions $\lambda_g, \lambda_{g,\text {low }} $ control the relative contribution of the two components. A low-mass tapering is introduced through an exponential cutoff governed by a smoothing parameter $\delta_m$~\cite{KAGRA:2021vkt}.
The secondary mass $m_{2}$ is drawn from a truncated power law conditional on the primary:
\begin{equation}
p(m_{2} | m_{1}, \Lambda_m) = \mathcal{P}(m_{2} | m_{\min}, m_{1}, \beta),
\end{equation}
with slope $\beta$ and lower bound $m_{\min}$.
Overall, the mass model is characterized by the following parameters:
\begin{align}
    \Lambda_m =
    \{
    \alpha, 
    \beta, 
    m_{\min}, 
    m_{\max},
    \delta_m,
    \mu_{g,{\rm low}}, 
    \sigma_{g,{\rm low}}, 
    \nonumber \\
    \mu_{g,{\rm high}},
    \sigma_{g,{\rm high}}, 
    \lambda_g,
    \lambda_{g,{\rm low}} 
    \}.
\end{align}
The parameters and model priors are reported in Table~\ref{tab:priors_masses}.

\begin{table}[t!]
\centering
\caption{Priors on the binary BH mass model hyperparameters adopted for all the analyses in this work. }
\label{tab:priors_masses}
\begin{tabularx}{\linewidth}{|X|l|l|}
\hline\hline
Power-Law + 2 Peaks mass model & Parameter & Prior range \\
\hline\hline
\multirow{11}{*}{Mass distribution} 
  & $\alpha$ & ${\cal U}[2, 5]$ \\
  & $\beta$ & ${\cal U}[-1, 5]$ \\
  & $m_{\min}$ & ${\cal U}[3, 8]\,M_\odot$ \\
  & $m_{\max}$ & ${\cal U}[70, 200]\,M_\odot$ \\
  & $\delta_m$ & ${\cal U}[0.1, 10]\,M_\odot$ \\
  & $\mu_{g, {\rm low}}$ & ${\cal U}[25, 40]\,M_\odot$ \\
  & $\sigma_{g, {\rm low}}$ & ${\cal U}[0.4, 10]\,M_\odot$ \\
  & $\mu_{g, {\rm high}}$ & ${\cal U}[40, 100]\,M_\odot$ \\
  & $\sigma_{g, {\rm high}}$ & ${\cal U}[0.4, 10]\,M_\odot$ \\
  & $\lambda_g$ & ${\cal U}[0, 1]$ \\
  & $\lambda_{g, {\rm low}}$ & ${\cal U}[0, 1]$ \\
\hline\hline
\end{tabularx}
\end{table}

\begin{figure*}[t]
    \centering
\includegraphics[width=1\textwidth]{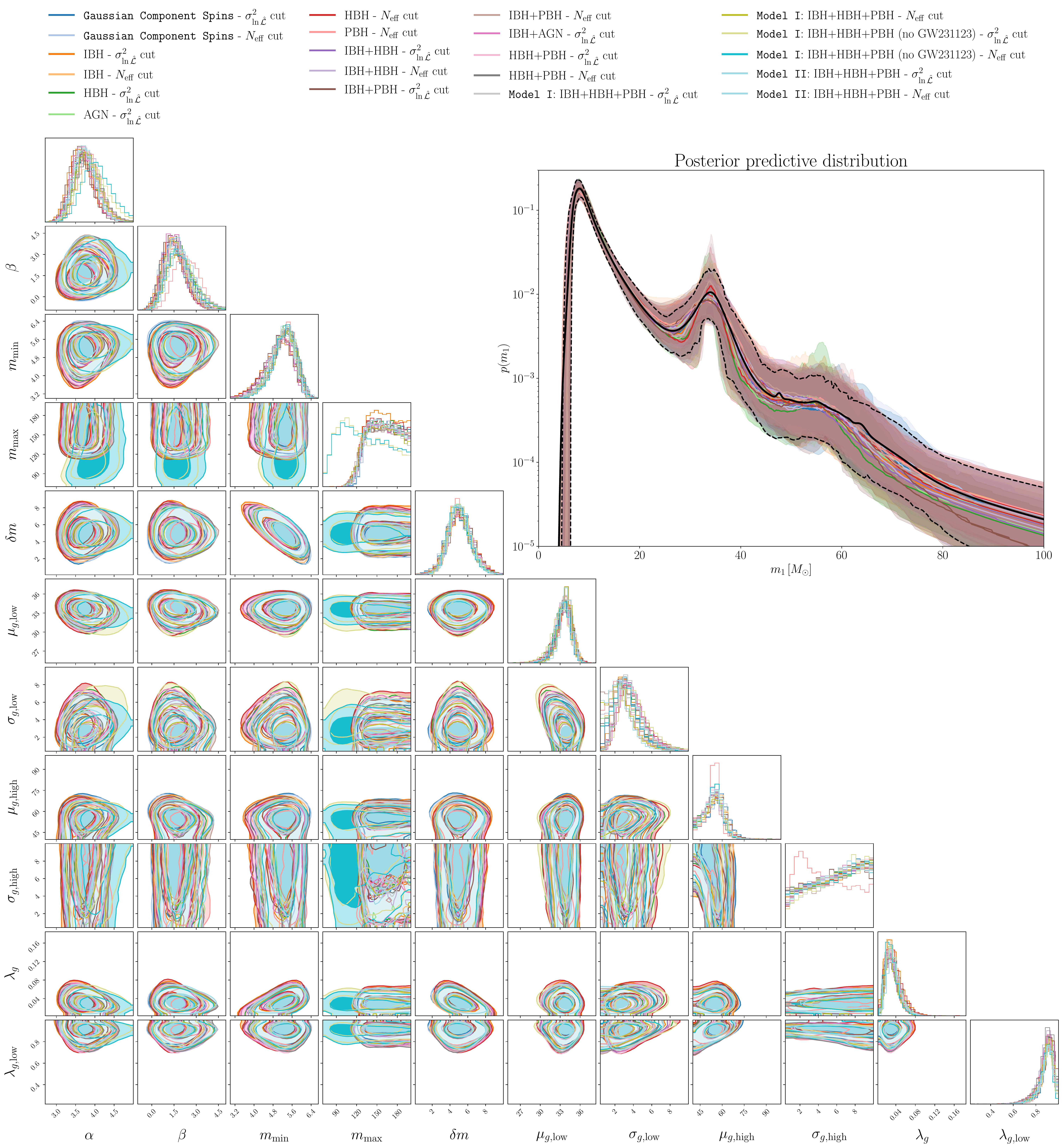}
    \caption{
    Posterior distributions of the mass model hyperparameters 
    inferred from the GWTC-4.0 catalog under different spin model assumptions. 
    We show results including and excluding GW231123, in order to test the 
    impact of this unique event on the inference. 
    Noticeable differences are: 
    (i) excluding GW231123 shifts 
    the preferred $m_{\rm max}$ to smaller values ($\sim 90\,M_\odot$);
    (ii) in the PBH-only scenario, the heavy bump is forced to lie more narrowly around $55\,M_\odot$. 
 }
    \label{fig:GWTC-4_allmasses}
\end{figure*}
In Fig.~\ref{fig:GWTC-4_allmasses}, we show the posterior distribution for the mass hyperparameters, obtained in all analyses performed in this work. We observe only statistically insignificant modifications in the PPD when varying across all the different spin model assumptions.

The only noticeable difference arises when removing GW231123 from the dataset. In this case $m_{\rm max}$ is 
shifted towards smaller values, down to around $90\,M_\odot$, consistent with the 
primary mass of GW190521. 
Secondly, we find that the PBH-only scenario forces the heavy bump to be more 
narrowly localized around $55\,M_\odot$. This behavior is likely related to the 
shape of the mass-spin correlation, which features a sharp transition from 
low-mass, low-spin systems to high-mass, high-spin systems. 
The presence of 
moderate spins at low masses requires this transition to occur at relatively 
intermediate/lower masses within the LVK range, corresponding to 
$z_{\rm cut-off} \lesssim 20$. However, this may lead to a tension with the 
moderately spinning events in the high-mass portion of the catalog, effectively 
forcing the preferred masses to be lighter. Overall, we stress that this model 
alone provides a poor fit to the full catalog (see also~\cite{Franciolini:2022tfm} for an analysis using GWTC-3 data).

\twocolumngrid
\bibliography{main}

\end{document}